\documentclass{aastex62}

\newcommand{\sfrac}[2]{\mathchoice
  {\kern0em\raise.5ex\hbox{\the\scriptfont0 #1}\kern-.15em/
   \kern-.15em\lower.25ex\hbox{\the\scriptfont0 #2}}
  {\kern0em\raise.5ex\hbox{\the\scriptfont0 #1}\kern-.15em/
   \kern-.15em\lower.25ex\hbox{\the\scriptfont0 #2}}
  {\kern0em\raise.5ex\hbox{\the\scriptscriptfont0 #1}\kern-.2em/
   \kern-.15em\lower.25ex\hbox{\the\scriptscriptfont0 #2}}
  {#1\!/#2}}

\newcommand{\myhalf}{\sfrac{1}{2}}

\newcommand{\eb}{{\bf{e}}}
\newcommand{\Ub}{{\bf{U}}}
\newcommand{\Ubt}{\widetilde{\Ub}}

\newcommand{\xb}{{\bf{x}}}

\newcommand{\dt}{\Delta t}

\newcommand{\etarho}{\eta_\rho}
\newcommand{\gammaonebar}{\overline{\Gamma}_1}
\newcommand{\Hnuc}{H_{\rm nuc}}
\newcommand{\omegadot}{\dot\omega}
\newcommand{\pred}{{\rm pred}}

\newcommand{\nph}{{n+\myhalf}}
\newcommand{\nmh}{{n-\myhalf}}

\newcommand{\uadvone}{\Ub^{\mathrm{ADV},\star}}
\newcommand{\uadvonedag}{\Ub^{\mathrm{ADV},\dagger,\star}}
\newcommand{\uadvtwo}{\Ub^{\mathrm{ADV}}}
\newcommand{\uadvtwodag}{\Ub^{\mathrm{ADV},\dagger}}

\usepackage{color}
\usepackage{url}

\usepackage{mathtools}

\usepackage{multirow}

\received{Aug 5, 2019}
\revised{XXX X, XXXX}
\accepted{XXX X, XXXX}
\submitjournal{ApJ}

\shorttitle{MAESTROeX Low Mach Number Astrophysics}
\shortauthors{Fan et al.}

\begin{document}

\title{MAESTROeX: A Massively Parallel Low Mach Number Astrophysical Solver}

\correspondingauthor{Duoming Fan; DFan@lbl.gov}

\author[0000-0002-3246-4315]{Duoming Fan}
\affil{Lawrence Berkeley National Laboratory \\
Center for Computational Sciences and Engineering \\
One Cyclotron Road, MS 50A-3111 \\
Berkeley, CA 94720, USA}

\author[0000-0003-1791-0265]{Andrew Nonaka}
\affil{Lawrence Berkeley National Laboratory \\
Center for Computational Sciences and Engineering \\
One Cyclotron Road, MS 50A-3111 \\
Berkeley, CA 94720, USA}

\author[0000-0003-2103-312X]{Ann S. Almgren}
\affil{Lawrence Berkeley National Laboratory \\
Center for Computational Sciences and Engineering \\
One Cyclotron Road, MS 50A-3111 \\
Berkeley, CA 94720, USA}

\author[0000-0002-1530-781X]{Alice Harpole}
\affil{Stony Brook University \\
Department of Physics and Astronomy \\
Stony Brook, NY 11794-3800, USA}

\author[0000-0001-8401-030X]{Michael Zingale}
\affil{Stony Brook University \\
Department of Physics and Astronomy \\
Stony Brook, NY 11794-3800, USA}



\begin{abstract}
We present MAESTROeX, a massively parallel solver for low Mach number astrophysical flows.
The underlying low Mach number equation set allows for efficient, long-time integration for highly subsonic flows compared to compressible approaches.
MAESTROeX is suitable for modeling full spherical stars as well as well as planar simulations of dynamics within localized regions of a star,
and can robustly handle several orders of magnitude of density and pressure stratification.
Previously, we have described the development of the predecessor of MAESTROeX, called MAESTRO, in a series of papers. 
Here, we present a new, greatly simplified temporal integration scheme that retains the same order of accuracy as our previous approaches. 
We also explore the use of alternative spatial mapping of the one-dimensional base state onto the full Cartesian grid.
The code leverages the new AMReX software framework for block-structured adaptive mesh refinement (AMR) applications, allowing for scalability to large fractions of leadership-class machines.
Using our previous studies on the convective phase of single-degenerate progenitor models of Type Ia supernovae as a guide, we characterize the performance of the code and validate the new algorithmic features.  Like MAESTRO, MAESTROeX is fully open source.
\end{abstract}



\section{Introduction} \label{sec:intro}
Many astrophysical flows are highly subsonic.  In this regime,
sound waves carry sufficiently little energy that they do not
significantly affect the convective dynamics of the system.  In many
of these flows, modeling long-time convective dynamics are of
interest, and numerical approaches based on compressible hydrodynamics
are intractable, even on modern supercomputers.  One approach to this
problem is to use low Mach number models.  In a low Mach number
approach, sound waves are eliminated from the governing equations
while retaining compressibilitiy effects due to, e.g., nuclear energy
release, stratification, compositional changes, and thermal diffusion.  When the Mach
number (the ratio of the characteristic fluid velocity over the
characteristic sound speed; Ma $= U/c$) is small, the resulting system
can be numerically integrated with much larger time steps than a
compressible model.  Specifically, the time step increase is at least
a factor of $\sim 1/{\rm Ma}$ larger.  Each time step is more
computationally expensive due to the presence of additional linear
solves, but for many problems of interest the overall gains in
efficiency can easily be an order of magnitude or more.

Low Mach number models have been developed for a variety of contexts
including combustion \citep{day2000numerical}, terrestrial atmospheric
modeling \citep{durran:1989,oneill:2014,duarte2015low}, and elastic
solids \citep{abbate2017all}.  For astrophysical applications, a
number of approaches to modeling low Mach number flows have been
developed in recent years.  One of the more similar approaches to ours is
\cite{Lin:2006}, however this approach is only first-order accurate and does
not account for atmospheric expansion.  There are also semi-implicit all-Mach
number solvers, where the Euler equations are split into an acoustic
part and an advective part
\citep{Kwatra2009,Degond2009,Cordier2012,Haack2012,Happenhofer2013,Chalons2016,Padioleau2019}.
The fast acoustic waves are then solved using implicit time
integration, while the slow material waves are solved explicitly.
Another approach is to use preconditioned all-Mach number solvers
\citep{Miczek2014,Barsukow2016}, where the numerical flux is
multiplied by a preconditioning matrix.  This reduces the stiffness of
the system at low Mach numbers, while retaining the correct scaling
behavior. In the reduced speed of sound technique (RSST) and related
methods, the speed of sound is artificially reduced by including a
suitable scaling factor in the continuity equation, reducing the
restriction on the size of the time step
\citep{Rempel2005,Hotta2012,Wang2015,Takeyama2017,Iijima2018}.
Finally, there are fully implicit time integration codes for the
compressible Euler equations
\citep{Viallet2011,kifonidis:2012,Viallet2015,Goffrey2016}.  The MUSIC code
uses fully implicit time integration for the compressible Euler
equations, which therefore allows for arbitrarily large time steps.

Previously, we developed the low Mach number astrophysical solver, MAESTRO.
MAESTRO is a block-structured, Cartesian grid finite-volume, adaptive mesh refinement (AMR)
code that has been successfully used for many years for a number of applications, detailed below.
Unlike several of the references above, MAESTRO is not an all-Mach solver, but is suitable for
flows where the Mach number is small ($\sim 0.1$ or smaller).
Furthermore, the low Mach number model in MAESTRO is specifically designed for, but not limited
to, astrophysical settings with significant atmospheric stratification.
This includes full spherical stars, as well as planar simulations of dynamics within localized
regions of a star.
The numerical methodology relies on an explicit Godunov approach for advection, a stiff ODE
solver for reactions (VODE, \citealt{vode}), and multigrid-based linear solvers for the
pressure-projection steps.  Thus, the time step is limited by an advective CFL constraint based on the 
fluid velocity, not the sound speed.
Central to the algorithm are time-varying, one-dimensional stratified background (or base) state
density and pressure fields that are held in hydrostatic equilibrium.
The base state density couples to the full state solution through buoyancy terms in the momentum equation,
and the base state pressure couples to the full state solution by constraining the evolution of the
thermodynamic variables to match this pressure.
The time-advancement strategy uses Strang splitting to integrate the thermodynamic variables, a
second-order projection method to integrate the velocity subject to a divergence constraint,
and a velocity splitting scheme that uses a radially-averaged velocity to hydrodynamically evolve the base state.
The original MAESTRO code was developed in the pure-Fortran 90 FBoxLib software framework, whereas
MAESTROeX is developed in the C++/F90 AMReX framework \citep{AMReX,AMReX_JOSS}.

The key numerical developments of the original MAESTRO algorithm are presented in a series of
papers which we refer to as Papers I-V:
\begin{itemize}
\item In Paper I \citep{MAESTRO_I}, we derive the low Mach number equation set for stratified
environments from the fully compressible equations.
\item In Paper II \citep{MAESTRO_II}, we incorporate the effects of atmospheric expansion
through the use of a time-dependent background state.
\item In Paper III \citep{MAESTRO_III}, we incorporate reactions and the associated coupling
to the hydrodynamics.
\item In Paper IV \citep{MAESTRO_IV}, we describe our treatment of spherical stars in a
three-dimensional Cartesian geometry.
\item In Paper V \citep{MAESTRO_V}, we describe the use of block-structured adaptive mesh
refinement to focus spatial resolution in regions of interest.
\end{itemize}

Since then, there have been many scientific investigations using MAESTRO, which have included additional algorithmic enhancements.  Topics include:
\begin{itemize}
\item The convective phase preceding Chandrasekhar mass models for type Ia supernovae \citep{MAESTRO_convection,MAESTRO_AMR,MAESTRO_CASTRO}.
\item Convection in massive stars \citep{Gilet:2013,gilkis:2016}.
\item Sub-Chandrasekhar white dwarfs \citep{subChandra_I,subChandra_II}.
\item Type I X-ray bursts \citep{XRB_I,XRB_II,XRB_III}.
\end{itemize}

In this paper, we present new algorithmic methodology that improves upon Paper V in a number of ways.
First, the overall temporal algorithm has been greatly simplified without compromising second-order accuracy.
The key design decisions were to eliminate the splitting of the velocity into average and perturbational components,
and also to replace the hydrodynamic evolution of the base state with a predictor-corrector approach.
Not only does this greatly simplify the dynamics of the base
state, but this treatment is more amenable to higher-order multiphysics coupling strategies
based on method-of-lines integration.
In particular, schemes based on deferred corrections \citep{dutt2000spectral} have been used to generate 
high-order temporal integrators for problems of reactive flow and low Mach number combustion \citep{pazner2016high,nonaka2018conservative}.
Second, we explore the effects of alternative spatial mapping routines for coupling the base state and the Cartesian grid state for spherical problems.
Finally, we examine the performance of our new MAESTROeX implementation in the new C++/F90 AMReX public software library \citep{AMReX,AMReX_JOSS}.
MAESTROeX uses MPI+OpenMP parallelism and scales well to over 10,000 MPI processes, with each MPI process supporting tens of threads.
The resulting code is publicly available on GitHub (\url{https://github.com/AMReX-Astro/MAESTROeX}),
uses the Starkiller-Astro microphysics libraries (\citealt{starkiller}, \url{https://github.com/starkiller-astro}) ,
as well as AMReX (\url{https://github.com/AMReX-Codes/amrex}).

The rest of this paper is organized as follows.
In Section \ref{sec:equations} we review our model for stratified low Mach number astrophysical flow.
In Section \ref{eq:algorithm} we present our numerical algorithm in detail, highlighting the new temporal integration scheme as well as spatial base state mapping options.
In Section \ref{sec:results} we validate our new approach and examine the performance of our algorithm on full spherical star problems used in previous scientific investigations.
We conclude in Section \ref{sec:conclusions}.

\section{Governing Equations}\label{sec:equations}
Low Mach number models for reacting flow were originally derived using asymptotic analysis
\citep{rehm1978equations,majda1985derivation} and used in terrestrial combustion applications
\citep{knio1999semi,day2000numerical}.  These models have been extended to nuclear flames
in astrophysical environments using adaptive algorithms in space and time \citep{Bell:2004}.
In Papers I-III, we extended this work and the atmospheric model by \citet{durran:1989} by deriving a model and algorithm suitable for stratified astrophysical flow.
We take the standard equations of reacting, compressible flow, and recast the equation
of state (EOS) as a divergence constraint on the velocity field.
The resulting model is a series of evolution equations for mass, momentum, and energy, subject
to an additional constraint on velocity.  The evolution equations are
\begin{eqnarray}
\frac{\partial(\rho X_k)}{\partial t} &=& -\nabla\cdot(\rho X_k\Ub) + \rho\omegadot_k,\label{eq:species}\\
\frac{\partial\Ub}{\partial t} &=& -\Ub\cdot\nabla\Ub  - \frac{\beta_0}{\rho}\nabla\left(\frac{\pi}{\beta_0}\right) - \frac{\rho-\rho_0}{\rho} g\eb_r,\label{eq:momentum}\\
\frac{\partial(\rho h)}{\partial t} &=& -\nabla\cdot(\rho h\Ub) + \frac{Dp_0}{Dt} + \rho\Hnuc.\label{eq:enthalpy}
\end{eqnarray}
Here $\rho$, $\Ub$, and $h$ are the mass density,
velocity and specific enthalpy, respectively, and
$X_k$ are the mass fractions of species $k$ with associated
production rate $\omegadot_k$ and energy release per time per unit mass $\Hnuc$.
The species are constrained such that $\sum_k X_k = 1$ giving $\rho = \sum_k (\rho X_k)$ and
\begin{equation}
\frac{\partial\rho}{\partial t} = -\nabla\cdot(\rho\Ub).
\end{equation}
The total pressure is decomposed into a one-dimensional hydrostatic base state
 pressure, $p_0 = p_0(r,t)$, and a dynamic pressure, $\pi = \pi(\xb,t)$, such that
$p = p_0 + \pi$ and $|\pi|/p_0 = \mathcal{O}({\rm Ma}^2)$ (we use $\xb$ to represent the Cartesian coordinate
directions of the full state and $r$ to represent the radial coordinate direction for the base state).
One way to mathematically think of the difference between $p_0$ and $\pi$ is that $\pi$ controls the velocity evolution 
in a way that forces the thermodynamic variables $(\rho,h,X_k)$ to evolve in a manner that is consistent with the EOS and $p_0$.

By comparing the momentum equation (\ref{eq:momentum}) to the momentum equation used in equation (2) in Paper V, we
note that we are using a formulation that enforces conservation of total energy in the
low Mach number system in the absence of external heating or viscous terms \citep{kleinpauluis,Vasil2013}.
We have previously validated this approach in modeling sub-Chandrasekhar white dwarfs using MAESTRO \citep{subChandra_II}.
We also define a one-dimensional base state density, $\rho_0 = \rho_0(r,t)$, that represents the lateral average (see Section \ref{Sec:Spatial}) of $\rho$ and is in hydrostatic equilibrium with $p_0$, i.e.,
\begin{equation}
\nabla p_0 = -\rho_0 g\eb_r, \label{eq:HSE}
\end{equation}
where $g=g(r,t)$ is the magnitude of the gravitational acceleration and $\eb_r$ is the unit vector in the outward radial direction.
Here $\beta_0$ is a density-like variable that carries background stratification, defined as
\begin{equation}
\beta_0(r,t) = \rho_0(0,t)\exp\left(\int_0^r\frac{1}{\gammaonebar p_0}\frac{\partial p_0}{\partial r'}dr'\right),
\end{equation}
where $\gammaonebar$ is the lateral average of $\Gamma_1 = d(\log p)/d(\log\rho) |_s$ (evaluated with entropy, $s$, constant).  We explored the effect of
replacing $\Gamma_1$ with $\gammaonebar$ as well as a correction term in paper III. 
Thermal diffusion is not discussed in this paper, but we have previously described the modifications to the original algorithm required
for implicit thermal diffusion in \cite{XRB_I}; inclusion of these effects in the new algorithm presented here is straightforward.

Mathematically, equations (\ref{eq:momentum})-(\ref{eq:enthalpy}) must still be closed by the EOS.
This is done by taking the Lagrangian derivative of the EOS for pressure as a function of the thermodynamic variables,
substituting in the equations of motion for mass and energy,
and requiring that the pressure is a prescribed function of altitude and time based on the hydrostatic equilibrium condition.
See Papers I and II for details of this derivation.
The resulting equation is a divergence constraint on the velocity field,
\begin{equation}
\nabla\cdot(\beta_0\Ub) = \beta_0\left(S - \frac{1}{\gammaonebar p_0}\frac{\partial p_0}{\partial t}\right).\label{eq:U divergence}
\end{equation}
The expansion term, $S$, incorporates local compressibility effects due to compositional changes and heat release from reactions,
\begin{equation}
S = -\sigma\sum_k\xi_k\omegadot_k + \frac{1}{\rho p_\rho}\sum_k p_{X_k}\omegadot_k + \sigma\Hnuc,\label{eq:S}
\end{equation}
where
$p_{X_k} \equiv \left. \partial p / \partial X_k \right|_{\rho,T,X_{j,j\ne k}}$,
$\xi_k \equiv \left. \partial h /\partial X_k \right |_{p,T,X_{j,j\ne k}}$,
$p_\rho \equiv \left.\partial p/\partial \rho \right |_{T, X_k}$, and
$\sigma \equiv p_T/(\rho c_p p_\rho)$, with $p_T \equiv \left. \partial p / \partial
T \right|_{\rho, X_k}$ and $c_p \equiv \left.  \partial h / \partial T
\right|_{p,X_k}$ is the specific heat at constant pressure.

To summarize, we model evolution equations for momentum, mass, and energy, equations (\ref{eq:momentum})-(\ref{eq:enthalpy}) subject to a divergence constraint on the velocity, equation (\ref{eq:U divergence}), and the hydrostatic equilibrium condition, equation (\ref{eq:HSE}).

\section{Numerical Algorithm}\label{eq:algorithm}
\subsection{Spatial Discretization}\label{Sec:Spatial}
The spatial discretization and adaptive mesh refinement methodology remains unchanged from Paper V.
We now summarize some of the key points here before describing the new temporal integrator in the next section.
We recommend the reader review Section 3 of Paper V for further details.

\begin{figure}[tb]
\centering
\includegraphics[height=2.0in]{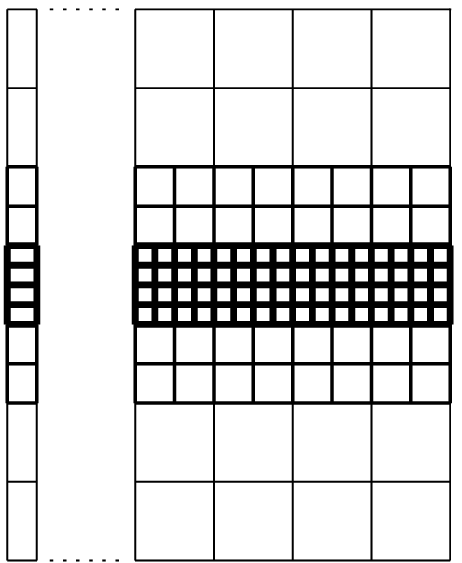} \hspace{0.5in}
\includegraphics[height=2.0in]{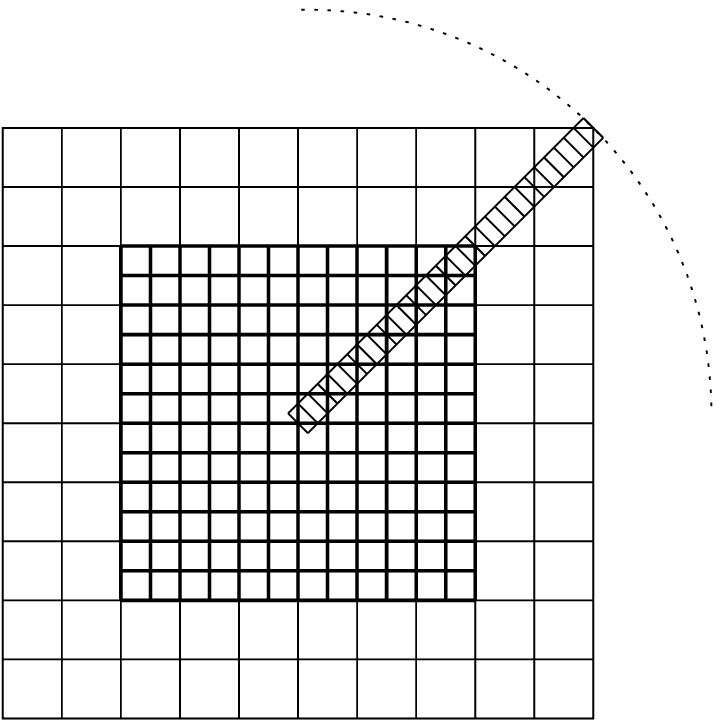}
\caption{\label{Fig:BaseGrid}  
(Left) For multi-level problems in planar geometry, we force a direct alignment
between the base state cell centers and the Cartesian grid cell centers by 
allowing the radial base state spacing to change with space and time.
(Right) For multi-level problems in spherical geometry, since there is no direct alignment
between the base state cell centers and the Cartesian grid cell centers, we choose to fix
the radial base state spacing across levels. Reprinted from Paper V \citep{MAESTRO_V}. }
\end{figure}

We shall refer to local atmospheric flows in two and three dimensions as problems in ``planar'' geometry, and full-star flows
in three dimensions as problems in ``spherical'' geometry.
The solution in both cases consists of the Cartesian grid solution
and the one-dimensional base state solution.
Figure \ref{Fig:BaseGrid} illustrates the relationship between the base state and the Cartesian grid state for both planar and spherical geometries in the presence of spatially adaptive grids. 
One of the key numerical modules is the ``lateral average'', which computes the average over a 
layer of constant radius of a Cartesian grid variable and stores the result in a one-dimensional base state array.
In planar geometries, this is a straightforward arithmetic average of values in cells at
a particular height since the base state cell centers are in alignment
with the Cartesian grid cell centers.
However for spherical problems, the procedure is much more complicated.
In Section 4 of Paper V, we describe how there is a finite, easily computable set of radii that any 
there-dimensional Cartesian cell-center can map to.  
Specifically, for every three-dimensional Cartesian cell, there exists an integer $m$ such that the distance
from the cell center to the center of the star is given by 
\begin{equation}
\hat{r}_m=\Delta x\sqrt{0.75+2m}.\label{eqn:radii}
\end{equation}
\begin{figure}[tb]
\centering
\includegraphics[height=2.5in]{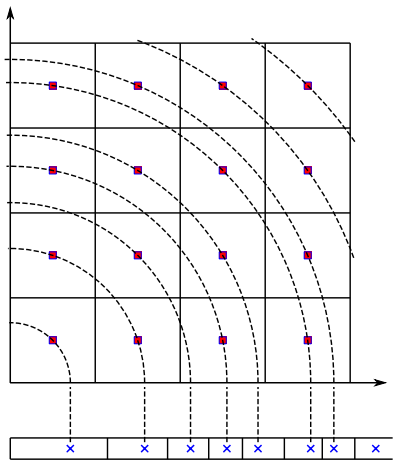}
\caption{\label{Fig:NewBaseGrid}  
A direct mapping between the base state cell centers (red squares) and the Cartesian grid cell centers (blue crosses) 
is enforced by computing the average of the grid cell centers that share the same radial distance from 
the center of the star.}
\end{figure}
Figure \ref{Fig:NewBaseGrid} is a two-dimensional illustration 
(two dimensions is chosen in the figure for ease of exposition; this mapping is only used for three-dimensional spherical stars) 
of the relationship between the Cartesian grid state and the one-dimensional base state array.
We compute the lateral average by first summing the values in all the cells associated with each radius,
dividing by the number of contributing cells to obtain the arithmetic average, and then use quadratic interpolation to map this data onto a one-dimensional base state.
Previously, for spherical problems MAESTRO only allowed for a base state with constant $\Delta r$ (typically equal to $\Delta x/5$).

The companion ``fill'' module maps a base state array onto the full Cartesian grid state.
For planar problems, direct injection can be used due to the perfect alignment of the base state and Cartesian grid state.
For spherical problems, quadratic interpolation of the base state is used to assign values to each Cartesian cell center.

In this paper we explore a new option to retain an irregularly-spaced base state to eliminate mapping errors from the ``fill'' module.
For the lateral average, as before we first sum the values in all the cells associated with each radius and divide
by the number of contributing cells to obtain the arithmetic average.  However we do not interpolate this onto a uniformly spaced
base state and retain the use of this irregularly-spaced base state.
The advantage with this approach is that the fill module does not require any interpolation.
A potential benefit to eliminating the mapping error is to consider a spherical star in hydrostatic equilibrium at rest.
In the absence of reactions, the star should remain at rest.
The buoyancy forcing term in the momentum equation contains $\rho-\rho_0$.  With the original scheme, interpolation errors 
in computing $\rho_0$ by averaging would cause artificial acceleration in the velocity field due to the interpolation error 
from the Cartesian grid to and from the radial base state.  By retaining the radial base state as an irregularly spaced 
array, the effects due to interpolation error are nearly eliminated, but not completely eliminated since there are still 
machine precision effects resulting from averaging a large number of numerical values.
We note that $\Delta r$ decreases as the base state moves further from the center of the star, 
which results in far more total cells in the irregularly-spaced array than the previous uniformly-spaced array. 

\subsection{Temporal Integration Scheme}\label{Sec:Temporal Integration Scheme}
We now describe the new temporal integration scheme, noting that it can be used for the original base state mapping 
(with uniform base state grid spacing) as well as the new irregularly spaced base state mapping.
Previously we adopted an approach where we split the velocity into a base state component, $w_0(r,t)$, 
and a local velocity $\Ubt(\xb,t)$, so that
\begin{equation}
\Ub = \Ubt(\xb,t) + w_0(r,t)\eb_r, \label{eq:velsplit}
\end{equation}
where $\eb_r$ is the normal vector in the outward radial direction.
We used $w_0$ to provide an estimate of the base state density evolution over a time step.
This resulted in some unnecessary complications to the temporal integration scheme including
base state advection modules for density, enthalpy, and velocity, as well as more 
cumbersome split velocity dynamics evolution equations.
Our new temporal integration scheme uses full velocities for scalar and velocity advection,
and only uses the above splitting to satisfy the velocity divergence constraint due to boundary considerations
at the edge of the star.
This results in a much simpler numerical scheme than the one from Paper V 
since we use the velocity directly rather than more complex terms involving the perturbational velocity.
Additionally, the new scheme uses a simpler predictor-corrector approach to the base state density and pressure that no 
longer requires evolution equations and numerical discretizations to update the base state, greatly 
simplifying the algorithm while retaining the same overall second-order accuracy in time.

At the beginning of each time step we have the cell-centered Cartesian grid state,
$(\Ub,\rho X_k,\rho h)^n$, and nodal Cartesian grid state, $\pi^{n-\myhalf}$, and base state $(\rho_0,p_0)^n$.
At any time, the associated density, composition, and enthalpy can be trivially computed using, e.g.,
\begin{equation}
\rho^n = \sum_k(\rho X_k)^n, \quad
X_k^n = (\rho X_k)^n / \rho^n, \quad
h^n = (\rho h)^n / \rho^n.
\end{equation}
Temperature is computed using the equation of state\footnote{As described in Paper V, for planar problems we compute temperature using $h$ instead of $p_0$, since we have successfully developed volume discrepancy schemes to effectively couple the enthalpy to the rest of the solution; see \cite{XRB_I}.  We are still exploring this option for spherical stars.}, e.g.,
$T = T(\rho,p_0,X_k)$, where $p_0$ has been mapped to the Cartesian grid using the fill module,
and ($\gammaonebar,\beta_0)$ are similarly computed from $(\rho,p_0,X_k)$;
see Appendix A of Paper I and Appendix C of Paper III for details on how $\beta_0$ is computed.

The overall flow of the algorithm begins with a second-order Strang splitting approach to integrate the advection-reaction system for 
the thermodynamic variables $(\rho X_k, \rho h)$, followed by a second-order projection methodology to integrate the velocities subject to a divergence constraint.  Within the thermodynamic variable update we use a predictor-corrector approach to achieve second-order accuracy in time.
To summarize:
\begin{itemize}
\item In {\bf Step 1} we react the thermodynamic variables over the first $\Delta t/2$ interval.
\item In {\bf Steps 2--4} we advect the thermodynamic variables over $\Delta t$.  Specifically, we compute an estimate for the expansion term, $S$, compute face-centered, time-centered velocities that satisfy the divergence constraint, and then advect the thermodynamic variables.
\item In {\bf Step 5} we react the thermodynamic variables over the second $\Delta t/2$ interval\footnote{After this step we could skip to the velocity advance in {\bf Steps 10--11}, however the overall scheme would be only first-order in time, so {\bf Steps 6-9} can be thought of as a trapezoidal corrector step.}.
\item In {\bf Steps 6--8} we redo the advection in {\bf Steps 2--4} but are able to use the trapezoidal rule to time-center certain quantities such as $S$, $\rho_0$, etc.
\item In {\bf Step 9} we redo the reactions from {\bf Step 5} beginning with the improved results from {\bf Steps 6--8}.
\item In {\bf Steps 10--11} we advect the velocity, and then project these velocities so they satisfy the divergence constraint while updating $\pi$.
\end{itemize}

There are a few key numerical modules we use in each time step.
\begin{itemize}
\item {\bf Average}$[\phi]\rightarrow[\overline\phi]$ computes the lateral average of a quantity over a layer at constant radius $r$, as described above in Section \ref{Sec:Spatial}.
\item {\bf Enforce HSE}$[\rho_0]\rightarrow[p_0]$ computes the base state pressure, $p_0$, from a base state density, $\rho_0$ by integrating the hydrostatic equilibrium condition in one dimension. 
This follows equation (A10) in Paper V, noting that for the irregularly spaced base state case, $\Delta r$ is not constant, where $\Delta r_{j+1/2} = r_{j+1} - r_j$ for cell face with index $j+1/2$.  
The base state pressure remains equal to a constant value at the location of a prescribed cutoff density outward for the entire simulation.
\item {\bf React State}$[(\rho X_k)^{\rm in},(\rho h)^{\rm in},p_0]\rightarrow[(\rho X_k)^{\rm out},(\rho h)^{\rm out},(\rho\dot\omega),(\rho\Hnuc)]$ 
uses VODE \citep{vode} to integrate the species and enthalpy due to reactions over $\Delta t/2$ by solving
\begin{equation}
\frac{dX_k}{dt} = \dot\omega_k(\rho,X_k,T); \qquad
\frac{dT}{dt} = \frac{1}{c_p}\left(-\sum_k\xi_k\dot\omega_k + H_{\rm nuc})\right)
\end{equation}
The inputs are the species, enthalpy, and base state pressure, and the outputs are the species, enthalpy, reaction rates, and nuclear energy generation rate.
See Paper III for details.
\end{itemize}
Each time step is constrained by the standard advective CFL condition,
\begin{equation}
\Delta t = \sigma^{\rm CFL} \min_i(\Delta x / U_i),
\end{equation}
where for our simulations we typically use $\sigma^{\rm CFL}\sim 0.7$ and the minimum is taken over all spatial directions over all cells.
There are additional constraints on the time step that are typically much less restrictive than the advective CFL including
the acceleration due to the buoyancy force (sometimes in effect when the velocity is approximately zero at the start of some simulations) 
and the local magnitude of the divergence constraint (to prevent too much mass evacuation from a cell in a time step); see Section 3.4 in Paper III for details.

In stratified low Mach number models, due to the extreme variation in density, the velocity can become very large in low density regions at the edge of the star.
These large velocities can severely affect the time step, so
throughout Papers II-V, we have employed two techniques to help control these dynamics without significantly 
affecting the dynamics in the convective region.
First, we use a cutoff density technique, where we hold the density constant outside a specified radius (typically near where the density is $\sim$4 orders of magnitude smaller than the largest densities in the simulation).
Second, we employ a sponge technique where we artificially damp the velocities near and beyond the cutoff region.
For more details, refer to Paper V and the previous references cited within.

Beginning with $(\Ub,\rho X_k,\rho h)^n$, $\pi^{n-\myhalf}$, and $(\rho_0,p_0)^n$,
the temporal integration scheme contains the following steps:
\begin{description}

\item[Step 1] {\em React the thermodynamic variables over the first $\Delta t / 2$ interval.}

Call {\bf React State}$[(\rho X_k)^n, (\rho h)^n, p_0^n] \rightarrow [(\rho X_k)^{(1)}, (\rho h)^{(1)}, (\rho \omegadot_k)^{(1)}, (\rho \Hnuc)^{(1)}]$.


\item[Step 2] {\em Compute the time-centered expansion term, $S^{\nph,\star}$.}

We compute an estimate for the time-centered expansion term in the velocity
divergence constraint.  Following \citet{Bell:2004}, we extrapolate
to the half-time using $S$ at the previous and current
time steps,
\begin{equation}
S^{\nph,\star} = S^n + \frac{\dt^n}{2} \left(\frac{S^n - S^{n-1}}{\dt^{n-1}}\right).
\end{equation}
Note that in the first time step we average $S^0$ and $S^1$ from the
initialization step.

\item[Step 3] {\em Construct a face-centered, time-centered advective velocity, $\uadvone$.}

The construction of face-centered time-centered states used to discretize the
advection terms for velocity, species, and enthalpy, are performed using
a standard multidimensional corner transport upwind approach
\citep{colella1990multidimensional,saltzman1994unsplit} with the piecewise-parabolic method (PPM)
one-dimensional tracing \citep{colella1984piecewise}.  The full details of this
Godunov advection approach for all steps in this algorithm are described 
in Appendix A of \cite{XRB_III}.
Here we use equation (\ref{eq:momentum}) to compute face-centered, time-centered velocities, $\uadvonedag$.
The $\dagger$ superscript refers to the fact that the predicted velocity field does not satisfy the divergence constraint,
\begin{equation}
\nabla \cdot \left(\beta_0^n \uadvone\right) = \beta_0^n \left[S^{\nph,\star} - \frac{1}{\gammaonebar^np_0^n}\left(\frac{\partial p_0}{\partial t}\right)^{n-\myhalf} \right].\label{eq:div1}
\end{equation}
 We project $\uadvonedag$ onto the space of velocities that satisfy the constraint to obtain $\uadvone$.
Each projection step in the algorithm involves the solution of a variable-coefficient Poisson solve using multigrid.
Note that we still employ velocity-splitting as described by equation (\ref{eq:velsplit}) for this step 
in order to enforce the appropriate behavior of the system near the edge of the star as determined by the cutoff density.
The details of this ``MAC'' projection are provided in Appendix \ref{Sec:Projection}.

\item[Step 4] {\em Advect the thermodynamic variables over a time interval of $\dt.$}

\begin{enumerate}
\renewcommand{\theenumi}{{\bf \Alph{enumi}}}

\item Update $(\rho X_k)$ using a discretized version of
\begin{equation}
\frac{\partial(\rho X_k)}{\partial t} = -\nabla\cdot(\rho X_k\Ub),
\end{equation}
where the reaction terms have been omitted since they were already 
accounted for in {\bf React State}.  The update consists of two steps:

\begin{enumerate}
\renewcommand{\labelenumii}{{\bf \roman{enumii}}.}

\item Compute the face-centered, time-centered species, $(\rho X_k)^{\nph,\pred,\star}$,
  for the conservative update of $(\rho X_k)^{(1)}$ using a Godunov approach \citep{XRB_III}.
  As described in Paper V, for robust numerical slope limiting we predict 
  $\rho'^n=\rho^n-\rho_0^n$ and $X_k^n$ to faces
  and here we spatially interpolate $\rho_0^n$ to faces to assemble the fluxes.

\item Evolve $(\rho X_k)^{(1)} \rightarrow (\rho X_k)^{(2),\star}$ using
\begin{equation}
(\rho X_k)^{(2),\star} = (\rho X_k)^{(1)}
  - \dt \left\{ \nabla \cdot \left[ \uadvone (\rho X_k)^{\nph,\pred,\star} \right] \right\},
\end{equation}

\end{enumerate}

\item Update $\rho_0$ by calling {\bf Average}$[\rho^{(2),\star}]\rightarrow[\rho_0^{n+1,\star}]$.

\item Update $p_0$ by calling {\bf Enforce HSE}$[\rho_0^{n+1,\star}] \rightarrow [p_0^{n+1,\star}]$.

\item Update the enthalpy using a discretized version of equation
\begin{equation}
\frac{\partial(\rho h)}{\partial t} = -\nabla\cdot(\rho h\Ub) + \frac{Dp_0}{Dt} + \rho\Hnuc,
\end{equation}
again omitting the reaction terms since we already accounted for
them in {\bf React State}.  This equation takes the form:
\begin{equation}
\frac{\partial (\rho h)}{\partial t}  = - \nabla \cdot (\rho h\Ub) + \frac{\partial p_0}{\partial t} + (\Ub \cdot \eb_r) \frac{\partial p_0}{\partial r}.
\end{equation}
For spherical geometry, we solve the analytically equivalent form,
\begin{equation}
\frac{\partial (\rho h)}{\partial t}  = - \nabla \cdot (\rho h\Ub) + \frac{\partial p_0}{\partial t} + \nabla \cdot (\Ub p_0) - p_0 \nabla \cdot \Ub.
\end{equation}
The update consists of two steps:

\begin{enumerate}
\renewcommand{\labelenumii}{{\bf \roman{enumii}}.}

\item Compute the face-centered, time-centered enthalpy, $(\rho h)^{\nph,\pred,\star},$
  for the conservative update of $(\rho h)^{(1)}$ using using a Godunov approach \citep{XRB_III}.
  As described in Paper V, for robust numerical slope limiting 
  we predict $(\rho h)'^n=(\rho h)^n-(\rho h)_0^n$ to faces,
  where $(\rho h)_0^n$ is obtained by calling {\bf Average}$[(\rho h)^n]\rightarrow[(\rho h)_0^n]$,
  and here we spatially interpolate $(\rho h)_0^n$ to faces to assemble the fluxes.

\item Evolve $(\rho h)^{(1)} \rightarrow (\rho h)^{(2),\star}$ using
\begin{equation}
(\rho h)^{(2),\star}
= (\rho h)^{(1)} - \dt \left\{ \nabla \cdot \left[ \uadvone (\rho h)^{\nph,\pred,\star} \right] \right\} + \Delta t\frac{Dp_0}{Dt}
\end{equation}
where here
\begin{equation}
\frac{Dp_0}{Dt} =
\begin{cases}
\frac{p_0^{n+1,*} - p_0^n}{\Delta t} + \left(\uadvone \cdot \eb_r\right) \left(\frac{\partial p_0}{\partial r} \right)^{n}& {\rm (planar)} \\
\frac{p_0^{n+1,*} - p_0^n}{\Delta t} + \left[ \nabla \cdot \left (\uadvone p_0^{\nph} \right ) - p_0^{\nph} \nabla \cdot \uadvone \right]& {\rm (spherical)}
\end{cases}
,
\end{equation}
and $p_0^\nph = (p_0^n+p_0^{n+1,*})/2$.
\end{enumerate}
\end{enumerate}

\item[Step 5] {\em React the thermodynamic variables over the second $\Delta t / 2$ interval.}

Call {\bf React State}$[ (\rho X_k)^{(2),\star}, (\rho h)^{(2),\star}, p_0^{n+1,\star}] 
\rightarrow 
[ (\rho X_k)^{n+1,\star}, (\rho h)^{n+1,\star}, (\rho \omegadot_k)^{n+1,\star}, (\rho \Hnuc)^{n+1,\star} ].$

\item[Step 6] {\em Compute the time-centered expansion term, $S^{\nph,\star}$.}

First, compute $S^{n+1,\star}$ with
\begin{equation}
S^{n+1,\star} =  \left(-\sigma  \sum_k  \xi_k  \omegadot_k  + \frac{1}{\rho p_\rho} \sum_k p_{X_k}  {\omegadot}_k + \sigma \Hnuc\right)^{n+1,\star}.
\end{equation}
  Then, define
\begin{equation}
 S^{\nph} = \frac{S^n + S^{n+1,\star}}{2},
\end{equation}

\item[Step 7] {\em Construct a face-centered, time-centered advective velocity, $\uadvtwo$.}

The procedure to construct $\uadvtwodag$ is identical to the Godunov procedure
for computing $\uadvonedag$ in {\bf Step 3}, but uses
the updated value $S^{\nph}$ rather than $S^{\nph,\star}$.
The $\dagger$ superscript refers to the fact that the predicted velocity field does not satisfy the divergence constraint,
\begin{equation}
\nabla \cdot \left(\beta_0^{\nph} \uadvtwo\right) =
\beta_0^{\nph} \left[S^{\nph} - \frac{1}{\gammaonebar^{\nph}p_0^{\nph}}\left(\frac{\partial p_0}{\partial t}\right)^{\nph}\right],\label{eq:div2}
\end{equation}
with
\begin{equation}
\beta_0^{\nph} = \frac{ \beta_0^n +  \beta_0^{n+1,\star} }{2}, \quad
\gammaonebar^{\nph} = \frac{ \gammaonebar^n +  \gammaonebar^{n+1,\star} }{2}.
\qquad
\end{equation}
we project $\uadvtwodag$ onto the space of velocities that satisfy the constraint to obtain $\uadvtwo$ using a MAC projection (see Appendix \ref{Sec:Projection}).

\item[Step 8] {\em Advect the thermodynamic variables over a time interval of $\dt.$}

\begin{enumerate}
\renewcommand{\theenumi}{{\bf \Alph{enumi}}}

\item Update $(\rho X_k)$.  This step is identical to {\bf Step 4A} except we use
  the updated values $\uadvtwo$ and $\rho_0^{n+1,\star}$ rather than
  $\uadvone$ and $\rho_0^n$.  In particular:

\begin{enumerate}
\renewcommand{\labelenumii}{{\bf \roman{enumii}}.}

\item Compute the face-centered, time-centered species, $(\rho X_k)^{\nph,\pred}$,
  for the conservative update of $(\rho X_k)^{(1)}$ using a Godunov approach \citep{XRB_III}.
  Again, we predict $\rho'^n=\rho^n-\rho_0^n$ and $X_k^n$ to faces
  but here we spatially interpolate $(\rho_0^n+\rho_0^{n+1,*})/2$ to faces to assemble the fluxes.

\item Evolve $(\rho X_k)^{(1)} \rightarrow (\rho X_k)^{(2)}$ using
\begin{equation}
(\rho X_k)^{(2)} = (\rho X_k)^{(1)}
- \dt \left\{ \nabla \cdot \left[\uadvtwo (\rho X_k)^{\nph,\pred} \right] \right\},
\end{equation}

\end{enumerate}

\item Update $\rho_0$ by calling {\bf Average}$[\rho^{(2)}]\rightarrow[\rho_0^{n+1}]$.

\item Update $p_0$ by calling {\bf Enforce HSE}$[\rho_0^{n+1}] \rightarrow [p_0^{n+1}]$.

\item Update the enthalpy.  This step is identical to {\bf Step 4D} except we use
  the updated values $\uadvtwo$, $\rho_0^{n+1}$, $(\rho h)_0^{n+1}$, and $p_0^{n+1}$
  rather than
  $\uadvone, \rho_0^n$, $(\rho h)_0^n$, and $p_0^n$.
  In particular:

\begin{enumerate}
\renewcommand{\labelenumii}{{\bf \roman{enumii}}.}

\item Compute the face-centered, time-centered enthalpy, $(\rho h)^{\nph,\pred},$
  for the conservative update of $(\rho h)^{(1)}$ using a Godunov approach \citep{XRB_III}.
  Again, we predict $(\rho h)'^n=(\rho h)^n-(\rho h)_0^n$ to faces
  but here we spatially interpolate $[(\rho h)_0^n)+(\rho h)_0^{n+1,*}]/2$ to faces to assemble the fluxes.

\item Evolve $(\rho h)^{(1)} \rightarrow (\rho h)^{(2)}$.
\begin{equation}
(\rho h)^{(2)}
= (\rho h)^{(1)} - \dt \left\{ \nabla \cdot \left[ \uadvtwo (\rho h)^{\nph,\pred} \right] \right\} + \Delta t\frac{Dp_0}{Dt}
\end{equation}
where here
\begin{equation}
\frac{Dp_0}{Dt} =
\begin{cases}
\frac{p_0^{n+1} - p_0^n}{\Delta t} + \left(\uadvtwo \cdot \eb_r\right) \left(\frac{\partial p_0}{\partial r} \right)^{n}& {\rm (planar)} \\
\frac{p_0^{n+1} - p_0^n}{\Delta t} + \left[ \nabla \cdot \left (\uadvtwo p_0^{\nph} \right ) - p_0^{\nph} \nabla \cdot \uadvtwo \right]& {\rm (spherical)}
\end{cases}
,
\end{equation}
and $p_0^\nph = (p_0^n+p_0^{n+1})/2$.
\end{enumerate}
\end{enumerate}

\item[Step 9] {\em React the thermodynamic variables over the second $\Delta t / 2$ interval.}

Call {\bf React State}$[(\rho X_k)^{(2)},(\rho h)^{(2)}, p_0^{n+1}] \rightarrow [(\rho X_k)^{n+1}, (\rho h)^{n+1}, (\rho \omegadot_k)^{n+1}, (\rho \Hnuc)^{n+1}].$

\item[Step 10] {\em Define the new-time expansion term, $S^{n+1}$.}

\begin{enumerate}
\renewcommand{\theenumi}{{\bf \Alph{enumi}}}
\item Define
\begin{equation}
  S^{n+1} =  \left(-\sigma  \sum_k  \xi_k \omegadot_k  + \sigma \Hnuc +
  \frac{1}{\rho p_\rho} \sum_k p_{X_k}  \omegadot_k\right)^{n+1}.
\end{equation}

\end{enumerate}

\item[Step 11] {\em Update the velocity}.

First, we compute the face-centered, time-centered velocities, $\Ub^{\nph,\pred}$
using a Godunov approach \citep{XRB_III}. Then, we update
the velocity field $\Ub^n$ to $\Ub^{n+1,\dagger}$ by discretizing
equation (\ref{eq:momentum}) as
\begin{equation}
\Ub^{n+1,\dagger}
= \Ub^n - \dt \left[\uadvtwo \cdot \nabla \Ub^{\nph,\pred} \right]
 - \dt \left[ \frac{\beta_0^\nph}{\rho^\nph} \nabla \left( \frac{\pi^\nmh}{\beta_0^\nmh} \right) + \frac{\left(\rho^\nph-\rho_0^\nph\right)}{\rho^\nph} g^{\nph} \eb_r \right],
\end{equation}
where
\begin{equation}
\rho^\nph = \frac{\rho^n + \rho^{n+1}}{2}, \qquad \rho_0^\nph = \frac{\rho_0^n + \rho_0^{n+1}}{2}.
\end{equation}
Again, the $\dagger$ superscript refers
to the fact that the updated velocity does not satisfy the divergence constraint,
\begin{equation}
\nabla \cdot \left(\beta_0^{n+1} \Ub^{n+1} \right) = \beta_0^{n+1} \left[ S^{n+1} - \frac{1}{\gammaonebar^{n+1}p_0^{n+1}}\left(\frac{\partial p_0}{\partial t}\right)^{\nph}\right].\label{eq:div3}
\end{equation}
We use an approximate projection to project $\Ub^{n+1,\dagger}$ onto the space of velocities that satisfy the constraint to obtain $\Ub^{n+1}$ using a ``nodal'' projection.
This projection necessarily differs from the MAC projection used in
{\bf Step 3} and {\bf Step 7} because the velocities in those steps are defined
on faces and $\Ub^{n+1}$ is defined at cell centers, requiring different divergence
and gradient operators.
Furthermore, as part of the nodal projection, we also define a nodal new-time perturbational pressure, $\pi^\nph$.
Refer to Appendix \ref{Sec:Projection} for more details.

\end{description}
This completes one step of the algorithm.

To initialize the simulation we use the same procedure described in Paper III.
At the beginning of each simulation, we define $(\Ub,\rho X_k,\rho h)$.
We set initial values for $\Ub, \rho X_k$, and $\rho h$ and perform a sequence of projections 
(to ensure the velocity field satisfies the divergence constraint) 
followed by a small number of steps of the temporal advancement scheme to iteratively 
find initial values for $\pi^{n-\myhalf}$ and $S^0$ and $S^1$ for use in the first time step.

Our approach to adaptive mesh refinement is algorithmically the same as the treatment described 
in Section 5 of Paper V; we refer the reader there for details.
MAESTROeX supports refinement ratios of 2 between levels.
We note that for spherical problems, AMR is only available for the case of a uniformly-spaced base state.

\section{Performance and Validation}\label{sec:results}

\subsection{Performance and Scaling}\label{sec:scaling}
We perform weak scaling tests for simulations of convection preceding ignition in a spherical, full-star Chandrasekhar mass white dwarf. 
The simulation setup remains the same as reported in Section 3 of \cite{MAESTRO_AMR} and originally used in \cite{MAESTRO_convection}, and thus we emphasize that these scaling tests are performed using meaningful, scientific calculations.
Here, we perform simulations using $256^3, 512^3, 768^3, 1024^3, 1280^3$, and $1536^3$ grid cells on a spatially uniform grid (no AMR).
We divide each simulation into $64^3$ grids, so these simulations contain between 64 grids ($256^3$) and 13,824 grids ($1536^3$).
These simulations were performed using the NERSC cori system on the Intel Xeon Phi (KNL) partition.
Each node contains 68 cores, each capable of supporting up to 4 hardware threads (i.e., a maximum of 272 hardware threads per node).
For these tests, we assign 4 MPI tasks to each node, and 16 OpenMP threads per MPI process.
Each MPI task is assigned to a single grid, so our tests use between 64 and 13,824 MPI processes (i.e., between 1,024 and 221,184 total OpenMP threads).
For $64^3$ grids we discovered that using more than 16 OpenMP threads did not decrease the wallclock time due to a lack of work available per grid; in principle one could use larger grids, fewer MPI processes, and more threads per MPI process to obtain a flatter weak scaling curve, however the overall wallclock time would increase except for extremely large numbers of MPI processes (beyond the range we tested here).
Thus, the more accurate measure of weak scaling is to consider the number of MPI processes, since the scaling plot would look virtually identical for larger thread counts.
Note that the largest simulation used roughly 36\% of the entire computational system.
\begin{figure}[htb]
\begin{center}
\includegraphics[width=3.0in]{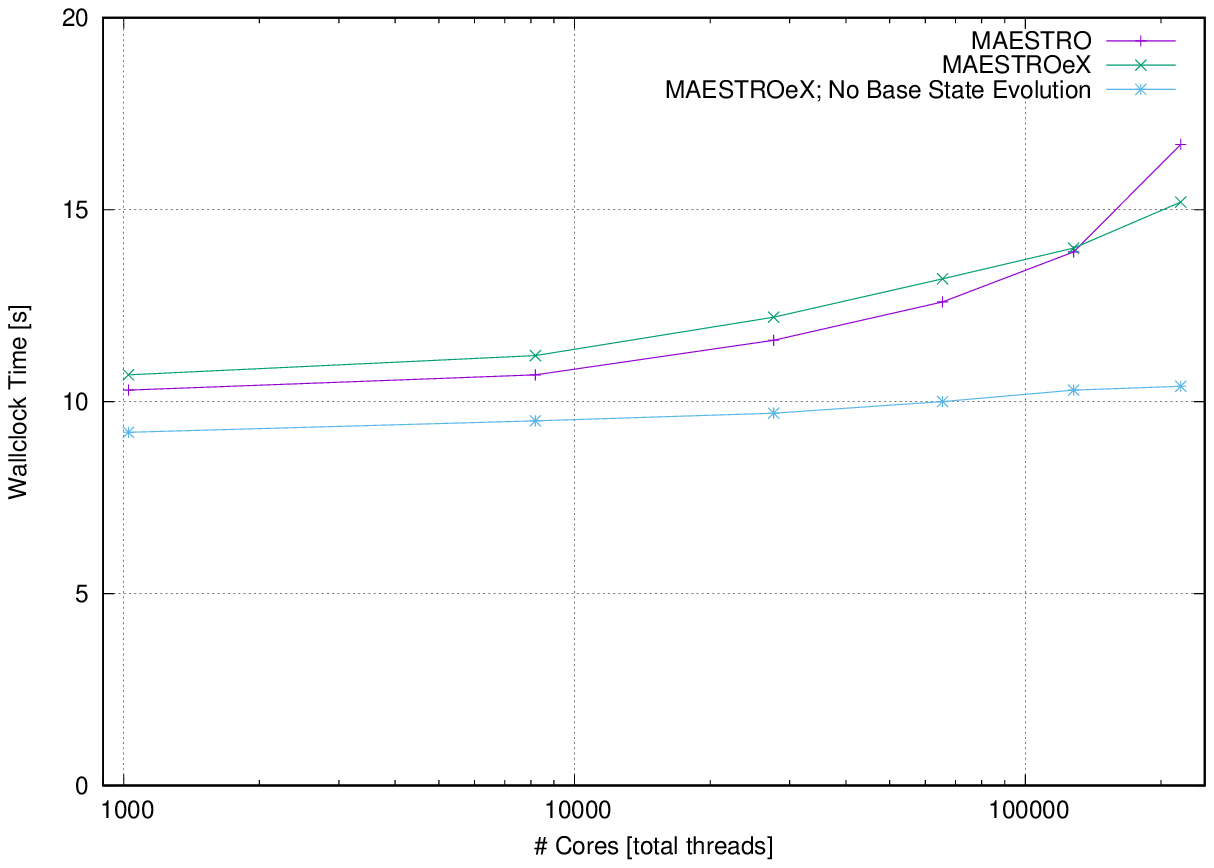} \hspace{0.5em}
\includegraphics[width=3.0in]{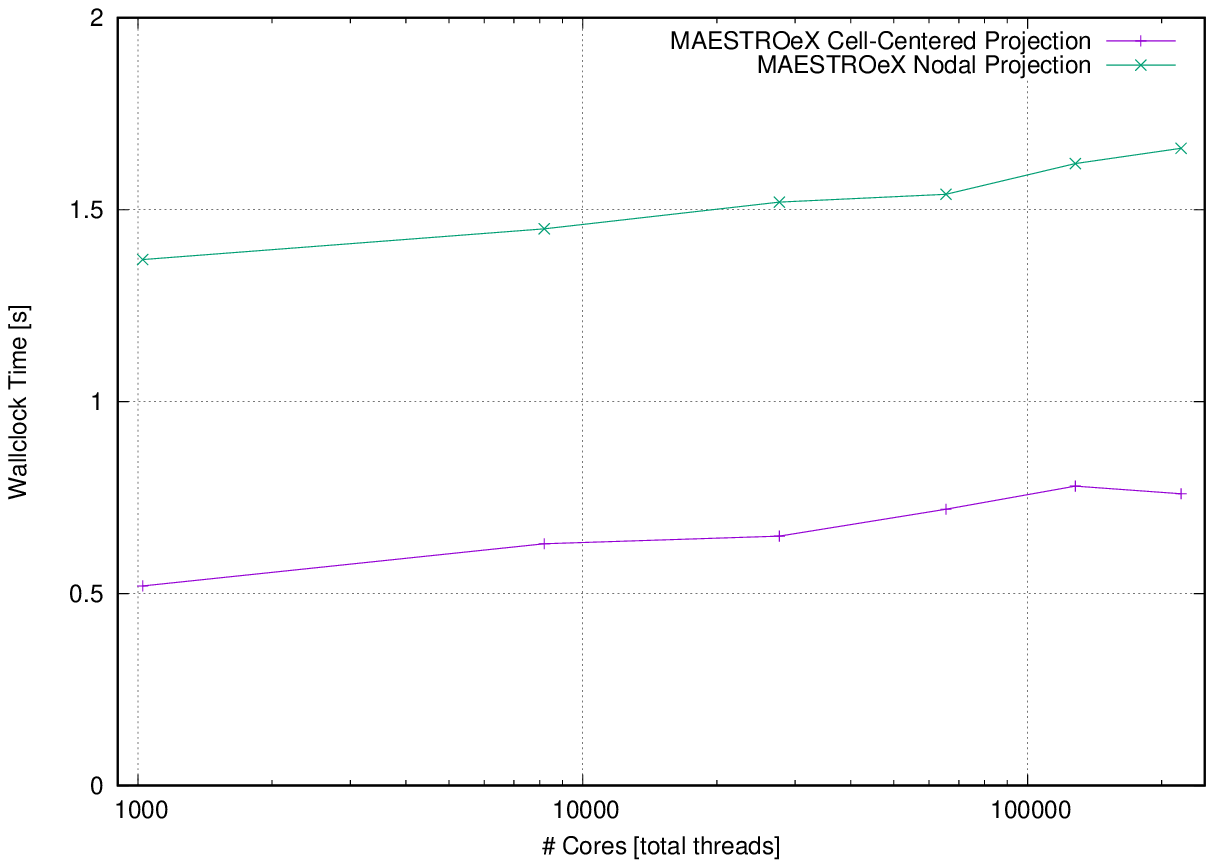}
\caption{\label{fig:scaling} (Left) Weak scaling results for a spherical, full-star white dwarf calculation using the original MAESTRO code, MAESTROeX, and MAESTROeX with base state evolution disabled.  Shown is the average wallclock time per time step.
(Right) Weak scaling results showing the average wallclock time per time step spent in the cell-centered and nodal linear solvers within a full time step of the aforementioned simulations.}
\end{center}
\end{figure}

In the left panel of Figure \ref{fig:scaling} we compare the wallclock time per time step as a function of total core count (in this case, the total number of OpenMP threads) for the original FBoxLib-based MAESTRO implementation to the AMReX MAESTROeX implementation.
These tests were performed using the original temporal integration strategy in \cite{MAESTRO_V}, noting that the new temporal integration with and without the irregular base state gives essentially the same results.
We also include a plot of MAESTROeX without base state evolution.
Comparing the original and new implementations, we see similar scaling results except for the largest simulation, where MAESTROeX performs better.
We see that the increase in wallclock time from the smallest to largest simulation is roughly 42\%.
We also note that without base state evolution, the code runs 14\% faster for small problems, and scales much better with wallclock time from the smallest to largest simulation increasing by only 13\%.
This is quite remarkable since there are 3 linear solves per time step (2 cell-centered Poisson solves used in the MAC projection, and a nodal Poisson solve used to compute the updated cell-centered velocities).
Contrary to our prior assumptions, the linear solves are not the primary scaling bottleneck in this code.
In the right panel of Figure \ref{fig:scaling}, we isolate the wallclock time required for these linear solves and see that (i) the linear solves only use 20-23\% of the total computational time, and (ii) the increase in the solver wallclock time from the smallest to largest simulation is only 28\%.
Further profiling reveals that the primary scaling bottleneck is the average operator.
The averaging operator requires collecting the sum of Cartesian data onto one-dimensional arrays holding every possible mapping radius.
This amounts to at least 24,384 double precision values (for the $256^3$ simulation) up to 883,584 values (for the $1536^3$ simulation).
The averaging operator requires a global sum reduction over all processors, and the communication of this data is the primary scaling bottleneck.
For the simulation with base state evolution, this averaging operator is only called once per time step (as opposed to 14 times per time step when base state evolution is included).
The difference in total wallclock times with and without base state evolution is almost entirely due to the averaging.
Note that as expected, advection, reactions, and calls to the equation of state scale almost perfectly, since there is only a single parallel communication call to fill ghost cells.

\subsection{White Dwarf Convection}\label{sec:whitedwarf}

\begin{figure}[htb]
\begin{center}
\includegraphics[width=2.75in]{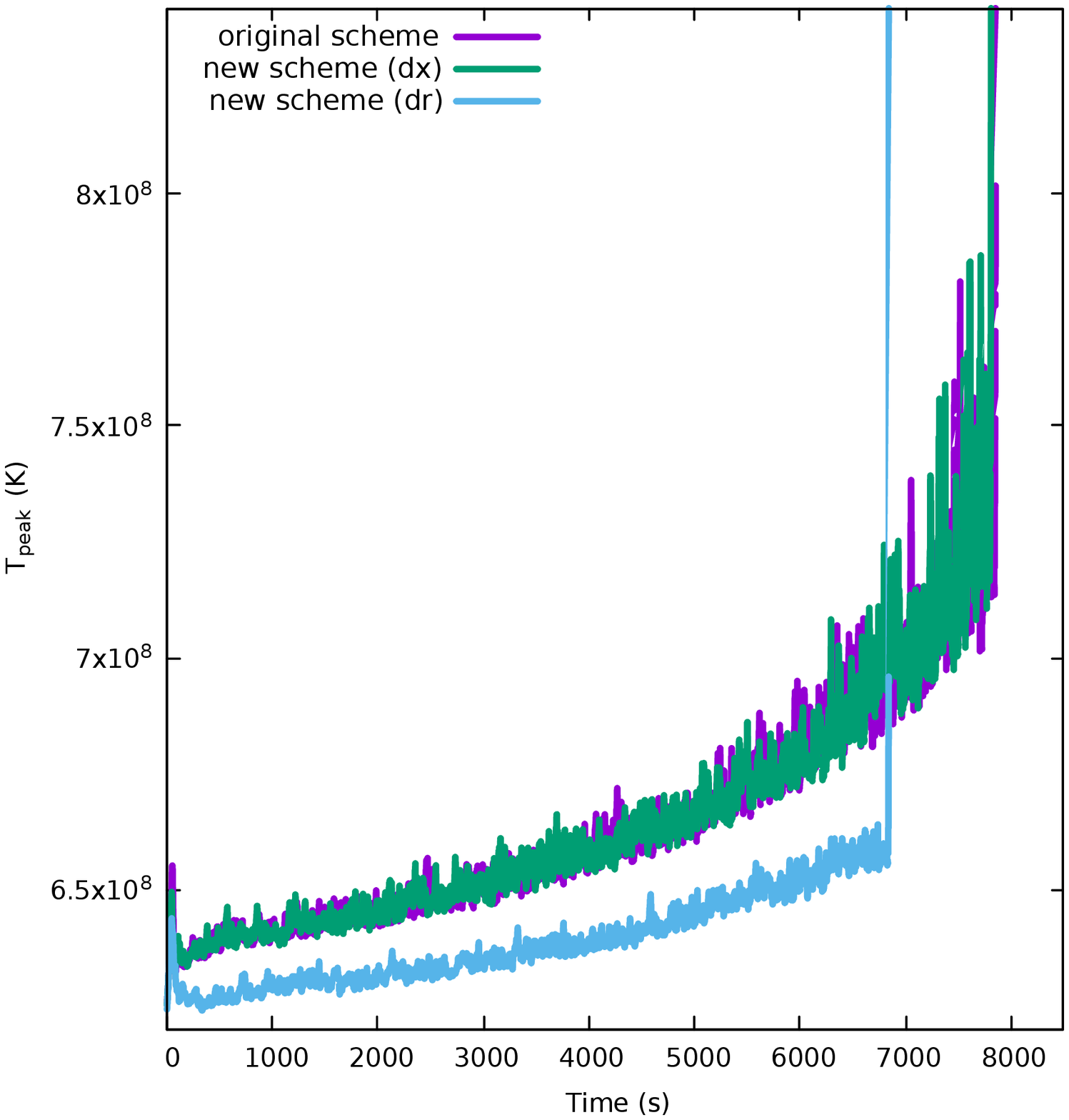}  \hspace{0.5in}
\includegraphics[width=2.75in]{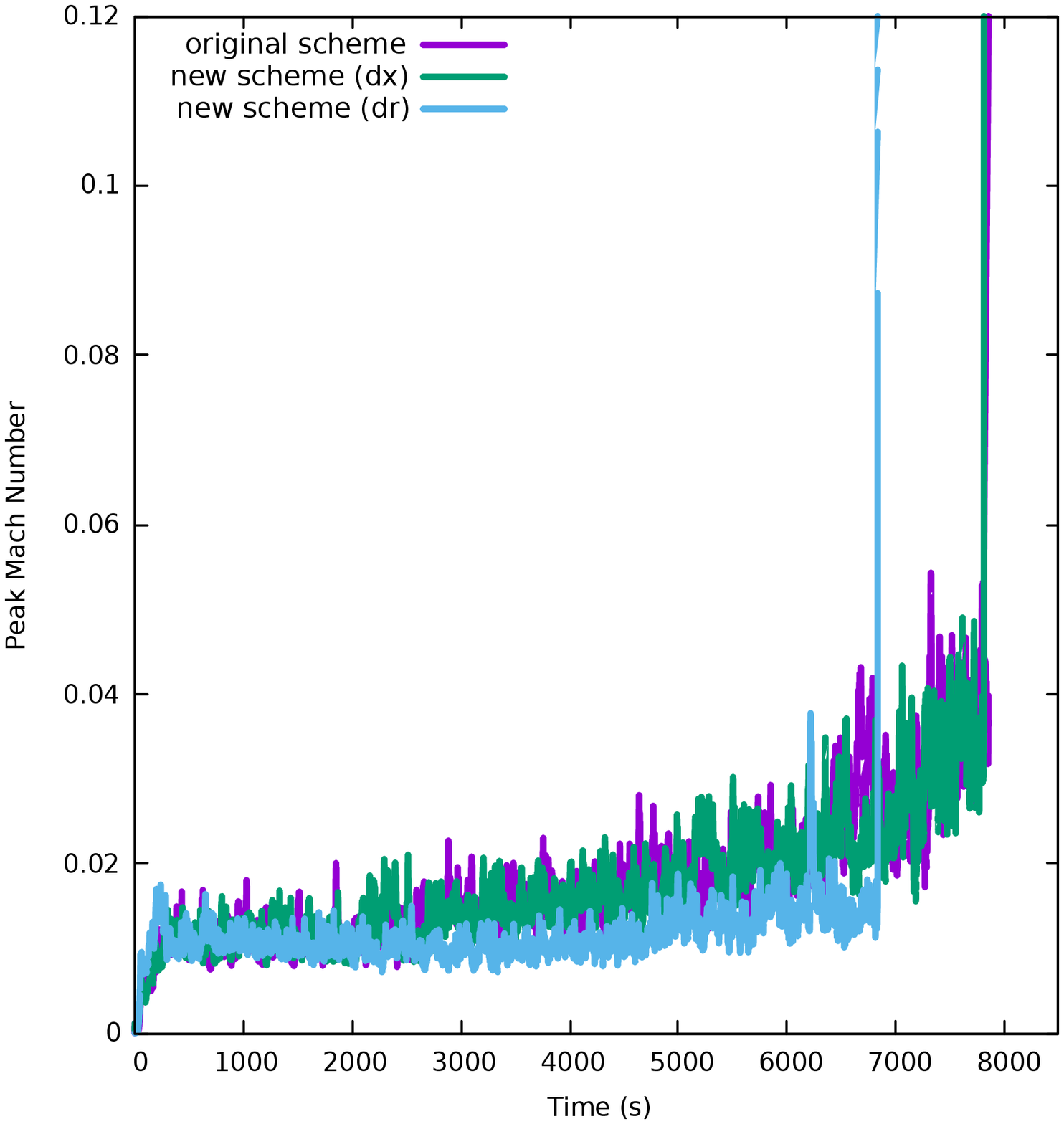}
\caption{\label{fig:wdconvect_256_maxvar} (left) Peak temperature, $T_{\text{peak}}$, and (right) peak Mach number
         in a white dwarf until time of ignition at resolution of $256^3$ for three different MAESTROeX algorithms.}
\end{center}
\end{figure}

To explore the accuracy of the new temporal algorithm, we now analyze in detail three-dimensional, full-star calculations of convection preceding ignition in a white dwarf. Again, we refer the reader to \cite{MAESTRO_AMR} and \cite{MAESTRO_convection} for setup details. We implement both uniformly- and irregularly-spaced base state with the new temporal algorithm, while only uniform base state spacing is used in the original algorithm. As in Section 3 of \cite{MAESTRO_AMR}, we choose the peak temperature and peak Mach number as the two diagnostics to compare the simulations. Figure \ref{fig:wdconvect_256_maxvar} shows the evolution of both peak temperature and peak Mach number until time of ignition on a single-level grid with resolution of $256^3$. The simulation using the new temporal scheme with uniformly-spaced base state gives the same qualitative results as the original scheme, and predicts a similar time of ignition ($t=7810$ s compared to $t=7850$ s for original algorithm). The simulation using the new temporal scheme with irregularly-spaced base state displays a slightly different peak temperature behavior during the initial transition period $t<150$ s, which results in the difference between the curves post transition. We strongly suspect that this is a result of using a different initial model file (the resolution near the center of the star is much coarser with the irregular spacing than the uniform spacing). Fortunately, the simulation with irregular base state spacing still follows the same trend as with uniform spacing, and the star is shown to ignite at an earlier time $t=6840$ s.

\begin{figure}[hbt]
\begin{center}
\includegraphics[width=3.25in]{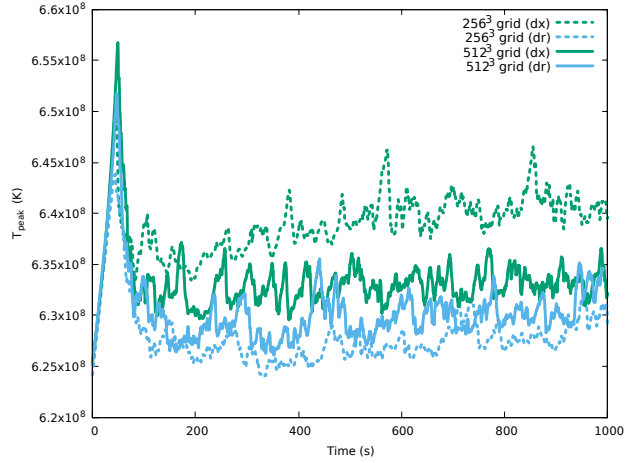}
\caption{\label{fig:wdconvect_Tmax} Peak temperature, $T_{\text{peak}}$  in a white dwarf at grid resolutions of
         $256^3$ (dotted line) and $512^3$ (solid line) until $t=1000$ for uniform (d$x$) and irregular (d$r$) base state spacing.
         Note that the irregularly-spaced solution agrees better
         with the uniformly-spaced solution as the resolution increases.}
\end{center}
\end{figure}

Figure \ref{fig:wdconvect_Tmax} shows the peak temperature evolution over the first 1000 s on two grids of differing resolutions, $256^3$ and $512^3$.
Limited allocations prevented us from running this simulation further.
As previously suspected, the simulation using irregularly-spaced base state agrees much closer with the results from using uniform spacing as the resolution increases.
This is most likely due to the increased resolution of the initial model, which more closely matches the uniformly-spaced counterpart.
This is especially important when computing the base state pressure from base state density, which is particularly sensitive to coarse resolution near the center of the star.

In terms of efficiency, all three simulations on $256^3$ single-level grid were run on Cori haswell with 64 processors and 8 threads per core and their run times were compared. As a result of simplifying the algorithm by eliminating the evolution equations for the base state density and pressure, the simulation using the new temporal algorithm took only 6.75 s per time step with uniformly-spaced base state, which is 13\% faster than the 7.77 s per time step when using the original scheme. However, we do observe a 25\% increase in run time of 9.72 s when using irregularly-spaced base state with the new algorithm. This can be explained by the irregularly-spaced base state array being much larger in size than its uniformly-spaced counterpart, and thus require additional communication and computation time. One possible strategy to significantly reduce the run time is to consider truncating the base state beyond the cutoff density radius.

\subsection{AMR Performance}

\begin{figure}[htb]
\begin{center}
\includegraphics[width=2.5in]{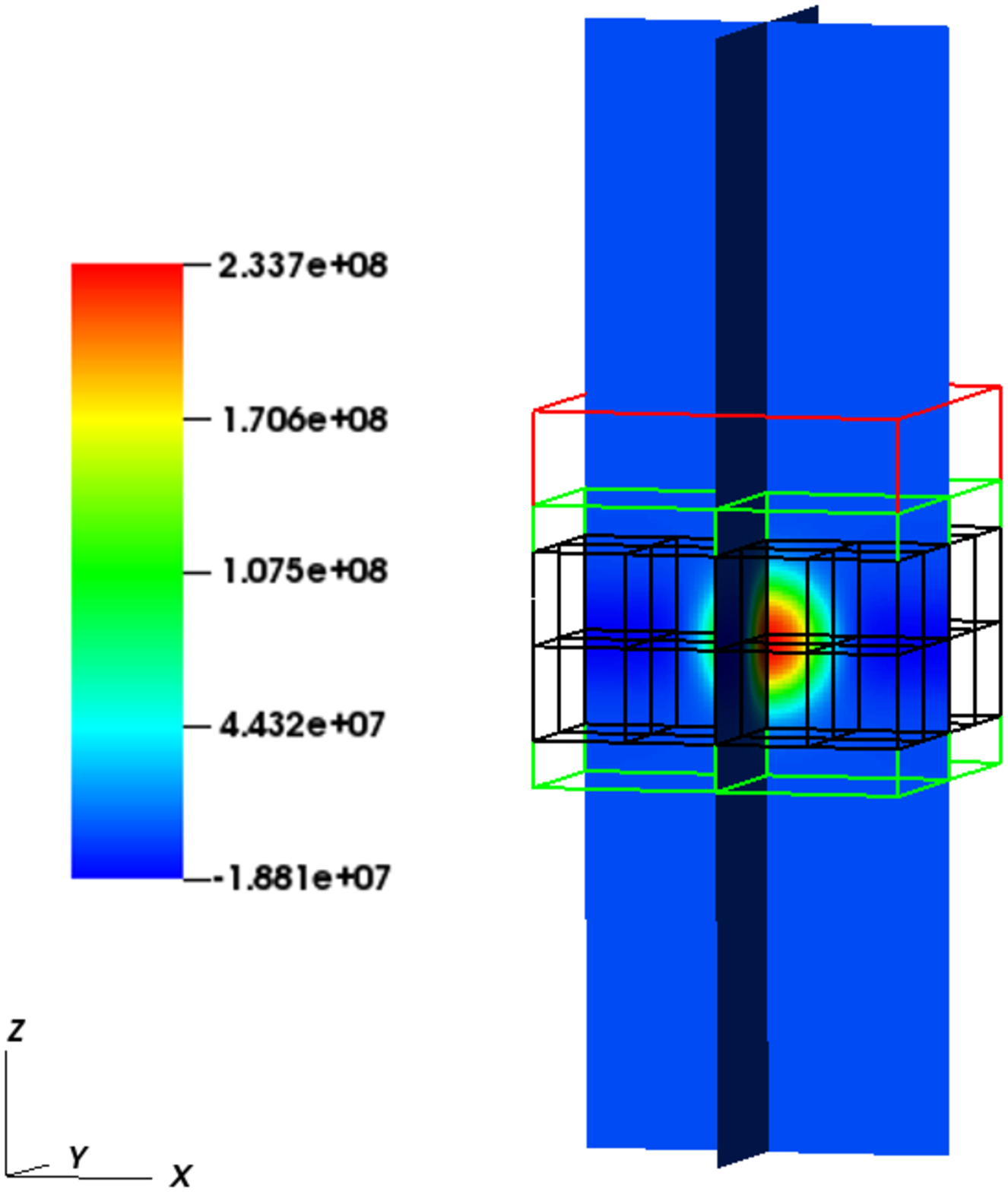} \hspace{2.5em}
\includegraphics[width=2.5in]{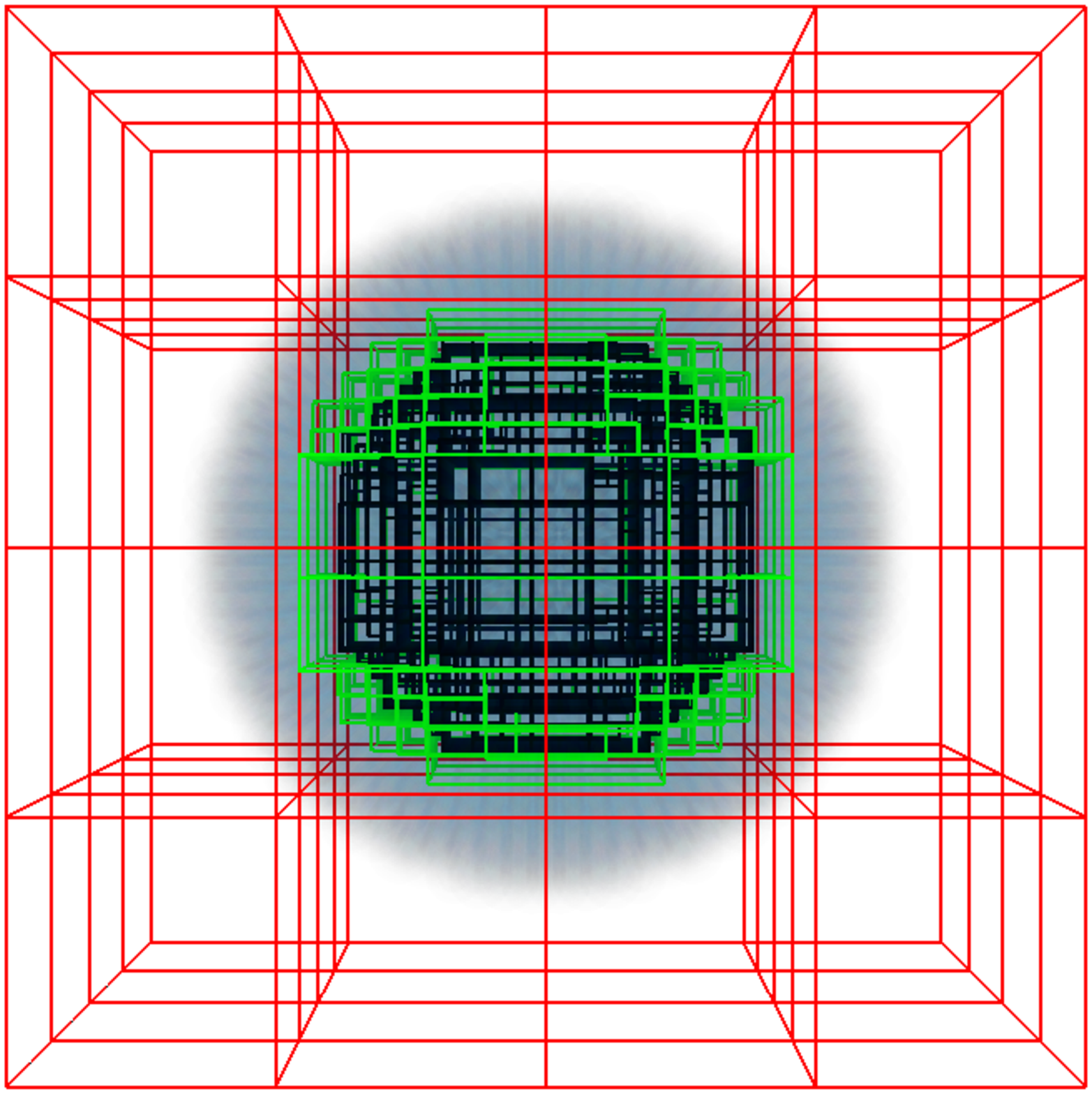}
\caption{\label{fig:amr_grids} Initial grid structures with two levels of refinement.
         The red, green, and black lines represent grids of increasing refinement.
         (Left) Profile of $T - \bar{T}$ for a hot bubble in a white dwarf environment.
         (Right) Region of star where $\rho\ge 10^5 \text{ g cm}^{-3}$ in a full white dwarf star simulation. }
\end{center}
\end{figure}

We now test the performance of MAESTROeX for adaptive, three-dimensional simulations to track localized regions of interest over time. Figure \ref{fig:amr_grids} illustrates the initial grid structures with two levels of refinement for both planar and spherical geometries where the grid is refined according to the temperature and density profiles, respectively. For each of the problems we tested, the single-level simulation was run using the original temporal scheme and the adaptive simulations using the original and new temporal algorithms. We want to show that the adaptive simulation can give similar results to the single-level simulation and in a more computationally efficient manner.

\begin{figure}[htb]
\begin{center}
\includegraphics[width=2.5in]{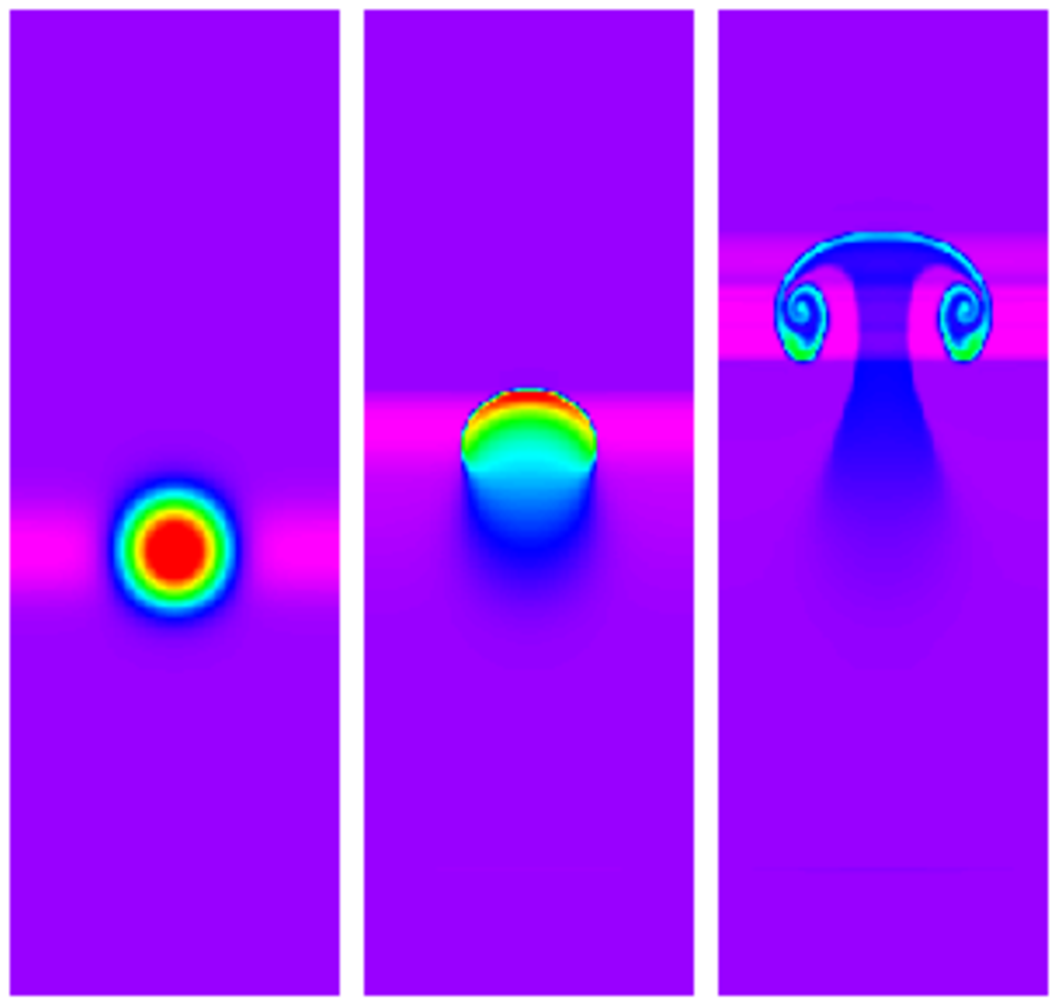} \hspace{2.5em}
\includegraphics[width=2.5in]{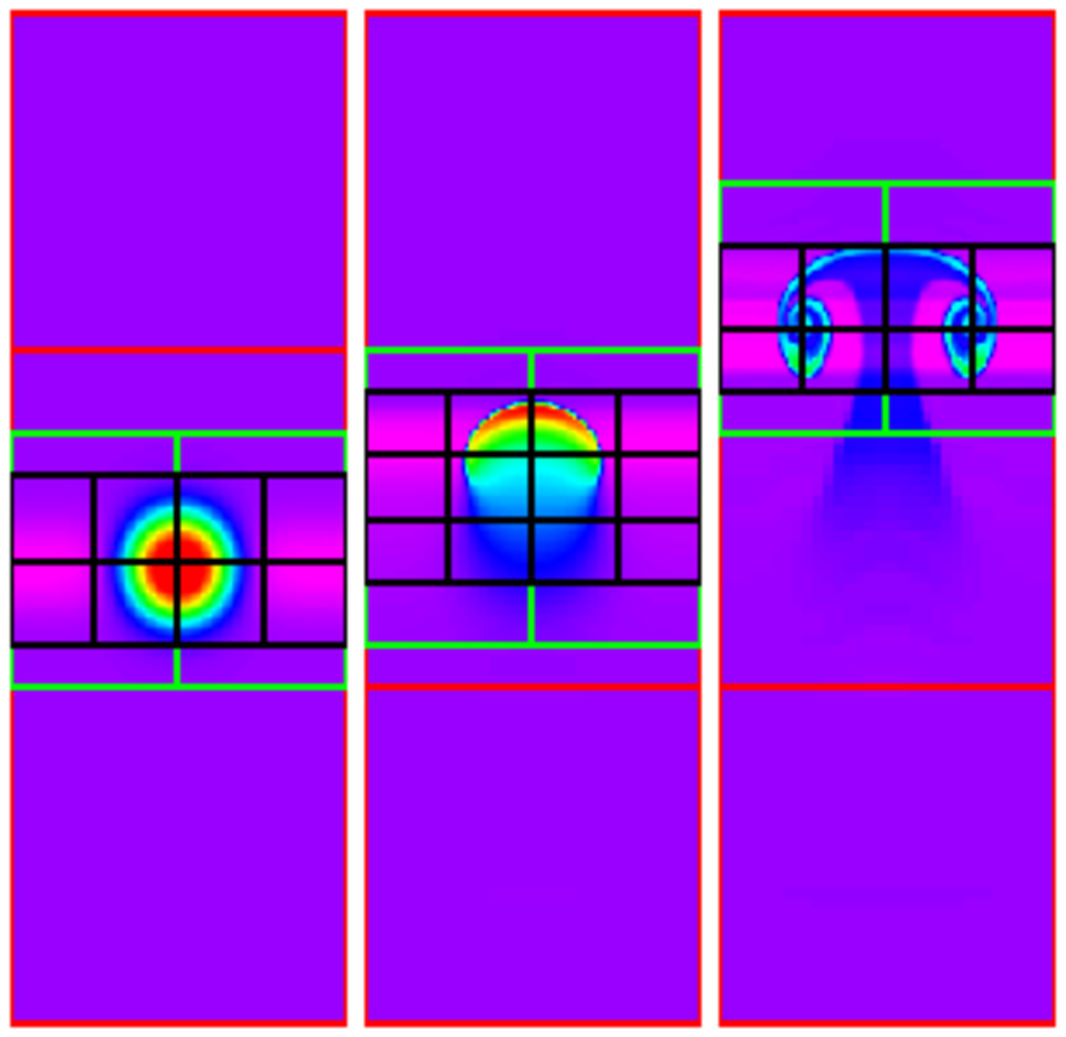}
\caption{\label{fig:bubble_results} Time-lapse cross-section of a hot bubble in a white dwarf environment at
         $t = 0$, 1.25, and 2.5 s for (left) single-level simulation, and
         (right) adaptive simulation at the same effective resolution.  The red, green, and black boxes indicate grids of increasing resolution}
\end{center}
\end{figure}

In the planar case, we use the same problem setup for a hot bubble rising in a white-dwarf environment as described in Section 6 of Paper V. Here we use a domain size of $3.6\times 10^7$ cm by $3.6\times 10^7$ cm by $2.88\times 10^8$ cm, and allow the grid structure to change with time. The single-level simulation at a resolution of $128^2 \times 1024$ was run on Cori haswell with 48 processors and took approximately 33.5 s per time step (averaged over 10 time steps) using either the original or new temporal algorithm. The adaptive simulation has a resolution of $32^2 \times 256$ at the coarsest level, resulting in the same effective resolution at the finest level as the single-level simulation. We tag cells that satisfy $T-\bar{T} > 3\times 10^7$ K as well as all cells at that height. The adaptive run took only 3.7 s per time step, and this 89\% decrease in runtime is mostly due to the fact that initially only 6.25\% of the cells (1,048,576 out of $128^2 \times 1024$ cells) are refined at the finest level. Figure \ref{fig:bubble_results} shows a series of planar slices of the temperature profile at time intervals of 1.25 s, and verifies that the adaptive simulation is able to capture the same dynamics as the single-level simulation at much lower computational cost.

\begin{figure}[htb]
\begin{center}
\includegraphics[width=4.0in]{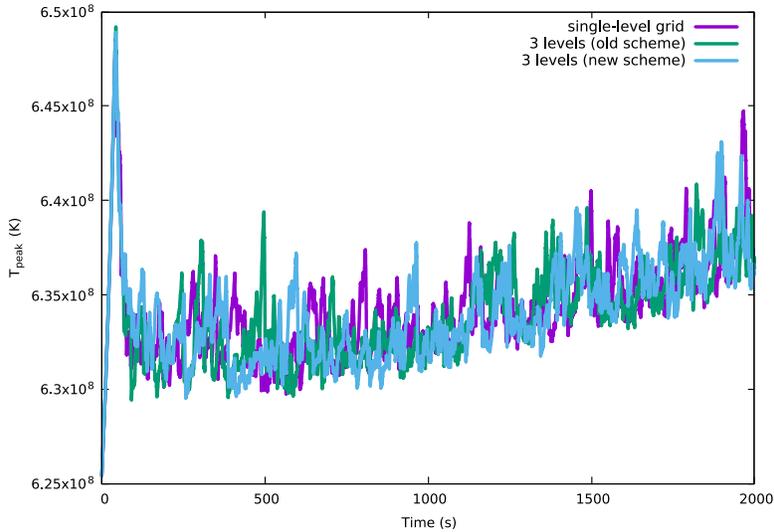}
\caption{\label{fig:wdconvect_amr_Tmax} Peak temperature, $T_{\text{peak}}$, in a white dwarf from $t=0$ to $2000$ s
         for grids with effective resolution of $512^3$. We can see that the adaptive grids with two levels of
         refinement give very similar solution trends compared to the single-level grid.}
\end{center}
\end{figure}

We continue to use the full-star white dwarf problem described in Section \ref{sec:whitedwarf} to test adaptive simulations on spherical geometry. The adaptive grid is refined twice by tagging the density at $\rho > 10^5$ g cm$^{-3}$ on the first level and $\rho > 10^8$ g cm$^{-3}$ on the second level. These tagging values have been shown to work well previously in Paper V, but we have found that the code may encounter numerical difficulties when the tagging values are too close to each other in subsequent levels of refinement.
The simulation on a single-level grid of $512^3$ resolution took 12.7 s per time step (again averaged over 10 time steps).  The adaptive grid has a resolution of $128^3$ at the coarsest level and an effective resolution of $512^3$ at the finest level. On this grid, 27.8\% of the cells (4,666,880 out of $256^3$ cells) are refined at the first level and 5.3\% (7,176,192 out of $512^3$ cells) at the second. The adaptive simulation took 5.61 s, resulting in more than a factor of 2 in speedup. Both simulations are computed to $t=2000$ s and we choose to use the peak temperature as the diagnostic to compare the results. Figure \ref{fig:wdconvect_amr_Tmax} shows the evolution of the peak temperature for all three runs and shows that the adaptive simulation gives the same qualitative result as the single-level simulation. We do not expect the curves to match up exactly because the governing equations are highly nonlinear, and slight differences in the solution caused by solver tolerance and discretization error can change the details of the results. Each simulation was run on Cori haswell with 512 processors and 4 threads per core.

\section{Conclusions and Future Work}\label{sec:conclusions}
We have developed a new temporal integrator and spatial mapping options into our low Mach number solver, MAESTROeX.
The new AMReX-enabled code scales well on large fractions of supercomputers with multicore architectures.
Future software enhancements will include GPU implementation.
In particular, the AMReX-based companion code, the compressible CASTRO code \citep{CASTRO}, has recently ported hydrodynamics and reactions to GPUs \citep{CASTRO_GPU}.
We plan to leverage the newly implemented mechanisms for offloading compute kernels to GPUs inside of the AMReX software library itself.
Our future scientific investigations include convection in massive rotating stars, \citep{heger2000presupernova}, the convective Urca process in white dwarfs \citep{willcox2016type}, solar physics \citep{wood2018self}, and magnetohydrodynamics \citep{wood2015three,wood2011sun}.
Our future algorithmic enhancements include more accurate and higher-order multiphysics coupling strategies based on spectral deferred corrections \citep{dutt2000spectral,bourlioux2003high}.
This framework has been successfully used in terrestrial combustion \citep{pazner2016high,nonaka2018conservative}

\acknowledgements

The work at LBNL was supported by the U.S. Department of Energy's
Scientific Discovery Through Advanced Computing (SciDAC) program under
contract No. DE-AC02-05CH11231.  The work at Stony Brook was supported
by DOE/Office of Nuclear Physics grant DE-FG02-87ER40317 and through
the SciDAC program DOE grant DE-SC0017955.  This research used
resources of the National Energy Research Scientific Computing Center
(NERSC), a U.S. Department of Energy Office of Science User Facility
operated under Contract No. DE-AC02-05CH11231.

\software{AMReX \citep{AMReX, AMReX_JOSS}, StarKiller Microphysics \citep{starkiller}}
\facilities{NERSC}

\appendix
\section{Projection Details}\label{Sec:Projection}
To enforce the divergence constraint in {\bf Step 3, 7}, and {\bf 11}, we use a projection method analogous to the methods originally developed for incompressible flow \citep{almgren1998conservative,bell1989second}.
The basic idea is to decompose the velocity field into a part that satisfies the divergence-satisfying component and a curl-free (gradient of a scalar field) component by solving a variable-coefficient Poisson equation for the scalar field.
The details for the MAC projection in {\bf Step 3} and {\bf Step 7} are given in Appendix B of Paper III.  The details of the nodal projection in {\bf Step 11} are given in Section 3.2 of Paper III.
We note that in the nodal projection, the gradient of the scalar field is used to update the perturbational pressure, $\pi$.

\begin{figure}[htb]
\begin{center}
\begin{tabular}{l c}
\multirow{4}{3.25in}[30mm]{ \includegraphics[width=3.0in]{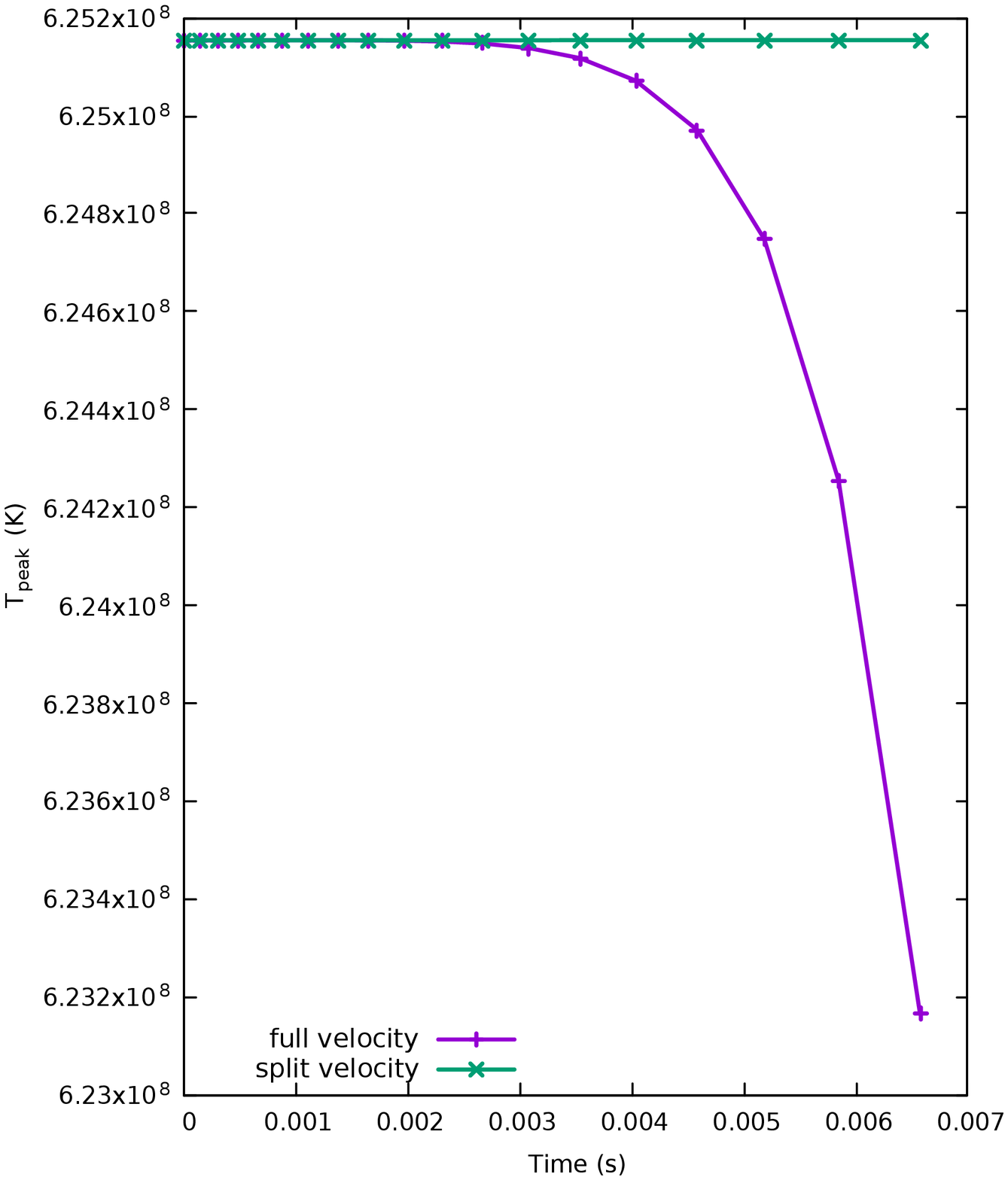} } & \multicolumn{1}{c}{\includegraphics[width=2.35in]{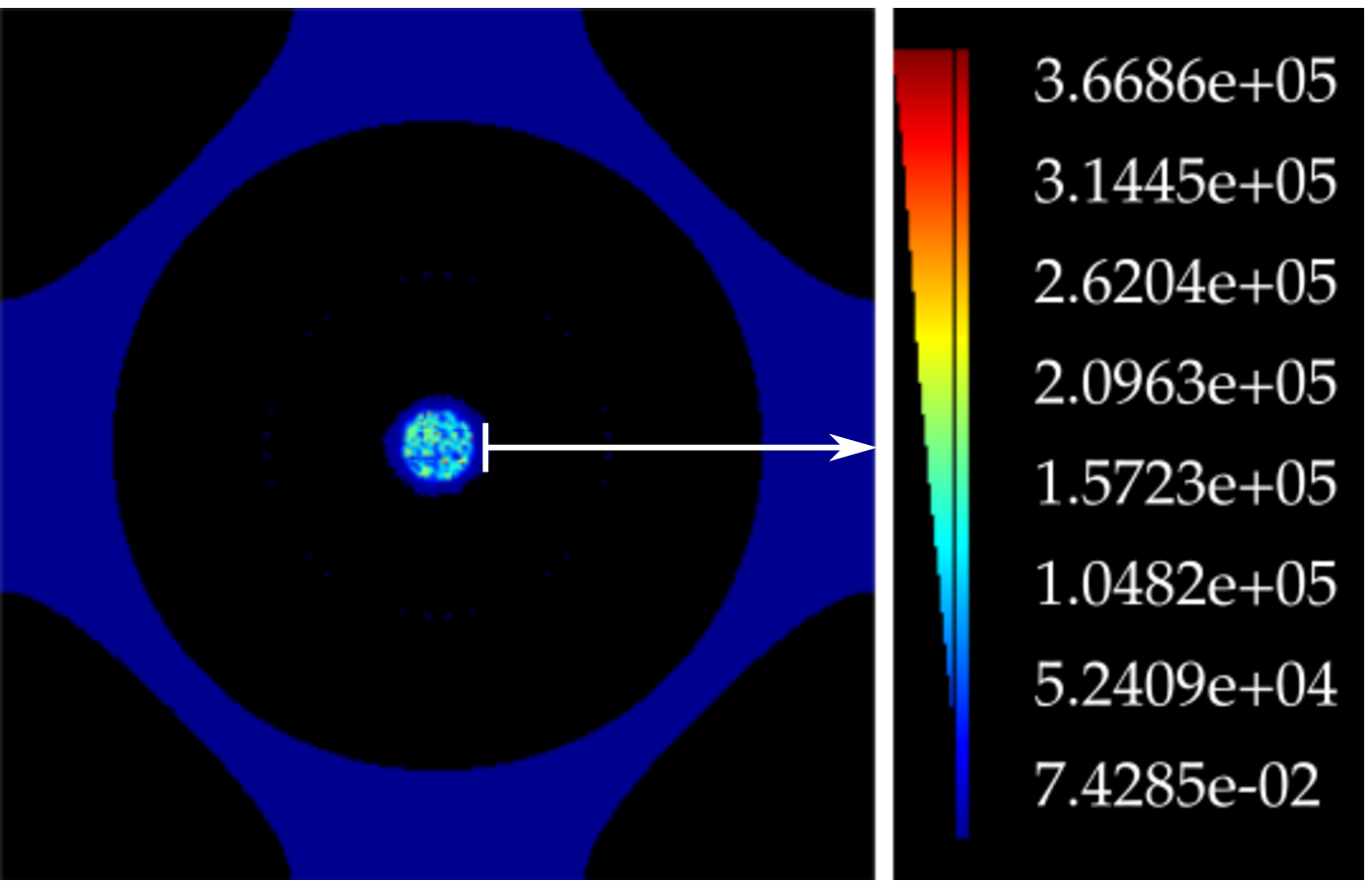}} \\
& \multicolumn{1}{c}{\begin{footnotesize} (a) Velocity magnitude, solved using full $\mathbf{U}$ \end{footnotesize}} \\[1.em]
& \multicolumn{1}{c}{\includegraphics[width=2.75in]{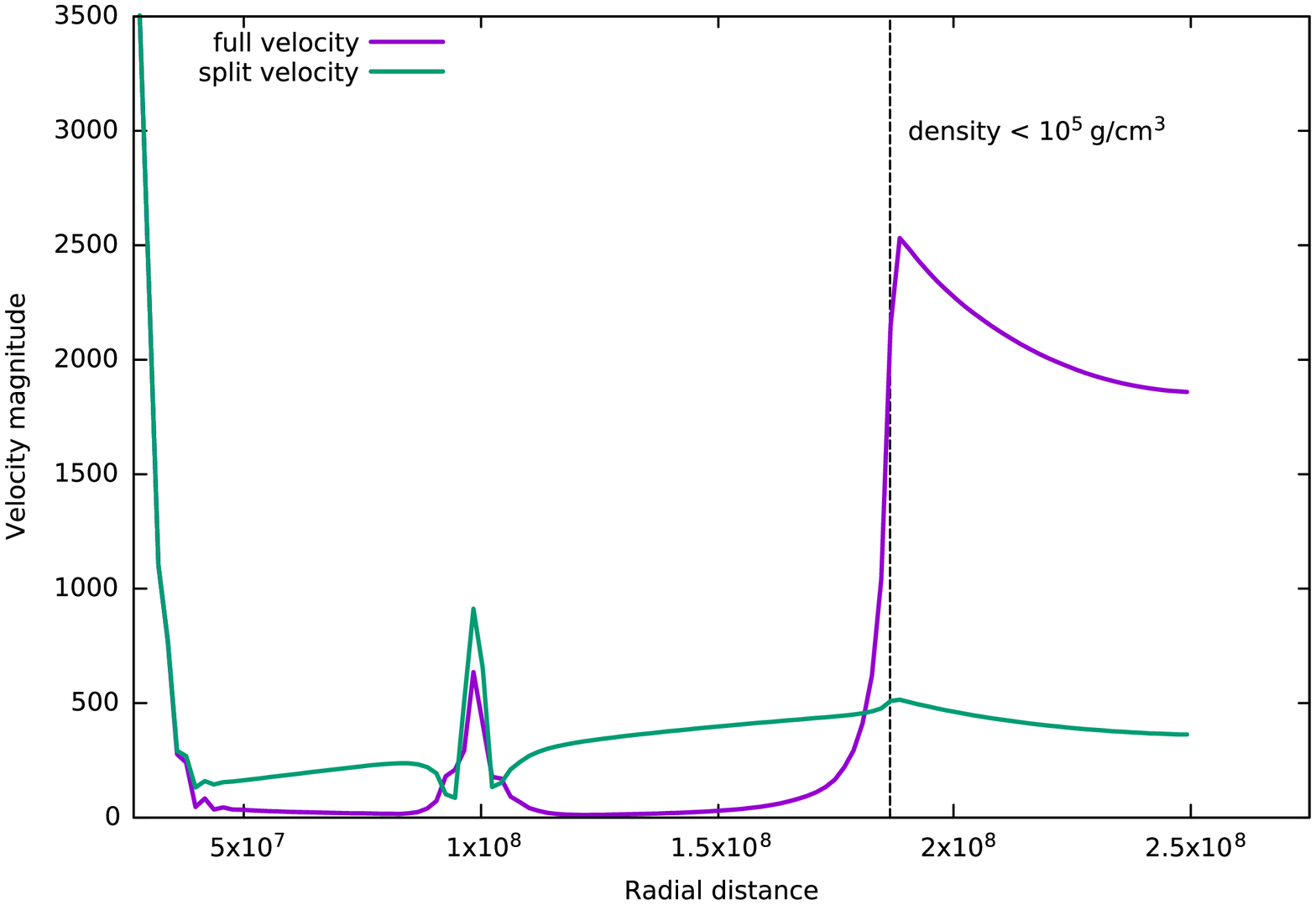}} \\ 
& \multicolumn{1}{c}{\begin{footnotesize} (b) Comparison along white line \end{footnotesize}} \\
\end{tabular}
\caption{\label{fig:wdconvect_splitU} When solving for the full velocity $\mathbf{U}$ in the nodal projection step instead of the split velocity 
  formulation, we illustrate
  (left) the peak temperature, $T_{\text{peak}}$ in a white dwarf at resolution of $256^3$ during the early evolution of the star and
  (right) the magnitude of velocity at early time. We observe much larger velocities outside the density cutoff region, $\rho < 10^5 \text{ g cm}^{-3}$, 
  that are not seen when we split the velocity dynamics.}
\end{center}
\end{figure}

Based on our past experience in the MAESTRO project, we have found it useful to split the velocity dynamics into a perturbational and base state velocity,
\begin{equation}
\Ub = \Ubt(\xb,t) + w_0(r,t)\eb_r,
\end{equation}
solve for each term separately, and immediately combine them to find a full velocity that satisfies the constraint.  We take that approach here, primarily because it allows us to enforce a boundary condition on $w_0$ at the edge of the star (i.e., the cutoff density location where we hold density constant).  Namely, to enforce that $r^2 w_0$ remain constant is difficult to do when solving for the full velocity. 
This is demonstrated in Figure \ref{fig:wdconvect_splitU} (right) where the velocity magnitude is observed to incorrectly increase outside the cutoff density radius when we solve for the full velocity in the nodal projection.
The resulting peak temperature also dips significantly as seen in Figure \ref{fig:wdconvect_splitU} (left), presumably because the dynamics of the overall expansion of the star are not being captured correctly.

In practice, we solve the constraint over the lateral average,
\begin{equation}
\nabla\cdot\left(\beta_0\Ubt\right) = \beta_0(S-\overline{S})
\end{equation}
and separately solve for $w_0$ using 
\begin{equation}
\nabla\cdot(\beta_0w_0\eb_r) = \beta_0\left(\overline{S} - \frac{1}{\gammaonebar p_0}\frac{\partial p_0}{\partial t}\right).
\end{equation}
To solve for $\Ubt$ we use a projection method, which involves the solution of a variable-coefficient Poisson solver to extract the curl-free component of the unprojected velocity, leaving a velocity field that satisfies the divergence constraint.
Note that MAESTRO contains alternate low Mach number formulations that conserve total energy in stratified systems, with minimal changes to the code (see Appendix A of \cite{subChandra_II} for details).
To find $w_0$ we integrate in one dimension using the procedure in Appendix B of Paper V, keeping in mind that the base state spacing ($\Delta r$) should be computed using the appropriate cell-edge and cell-center locations when using irregularly-spaced base state.
Note that in this approach, we estimated the time-derivative of the pressure in part by examining how laterally averaged $\rho'=\rho-\rho_0$ changed over time (quantified by $\eta_\rho = \overline{\rho'\Ub\cdot\eb_r}$).  After evolving the species ({\bf Steps 4A/8A}), we compute this term as reported in Paper V. For example, after {\bf Step 8A}, we define a radial cell-centered $\etarho^{\nph}$, 

\begin{description}
\item For planar geometry, $\etarho = \overline{\rho'(\Ub\cdot\eb_r)}$,
\begin{equation}
 \etarho^{\nph} =  {\rm {\bf Average}} \sum_k \left[ \left(\uadvtwo \cdot \eb_r \right) (\rho X_k)^{\nph,\pred} \right]
\end{equation}
\item For spherical geometry, first construct 
$\etarho^{{\rm cart},\nph} = [\rho'(\Ub\cdot\eb_r)]^{\nph}$ on Cartesian cell centers using:
\begin{equation}
\etarho^{{\rm cart},\nph} = \left[\left(\frac{\rho^{(1)}+\rho^{(2)}}{2}\right)-\left(\frac{\rho_0^n+\rho_0^{n+1}}{2}\right)\right] \cdot \left( \uadvtwo \cdot \eb_r \right).
\end{equation}
Then,
\begin{equation}
\etarho^{\nph} = {\rm {\bf Average}}\left(\etarho^{{\rm cart},\nph}\right).
\end{equation}
\end{description}

\bibliographystyle{aasjournal.bst}
\bibliography{references}

\end{document}